\documentclass[preprintnumbers,showpacs,amsmath,amssymb,floatfix,9pt,prd,superscriptaddress,nofootinbib]{revtex4}
\usepackage{graphicx}
\usepackage{latexsym}
\usepackage{epsfig}
\usepackage{amssymb}
\usepackage{color}
\usepackage{float}
\usepackage{subfigure}
\usepackage{hyperref}
\begin{document}
\title{Charged anisotropic strange stars in Brans-Dicke gravity with a massive scalar field through embedding approach}   

\author{S. K. Maurya}
\email{sunil@unizwa.edu.om} 
\affiliation{Department of Mathematical and Physical Sciences,
College of Arts and Science, University of Nizwa, Nizwa, Sultanate of Oman}

\author{Ksh. Newton Singh}
\email{ntnphy@gmail.coml
 } \affiliation{Department of Physics, National Defence Academy, Khadakwasla, Pune-411023, India.}

\author{M. Govender}
\email{megandhreng@dut.ac.za} \affiliation{Department of Mathematics, Durban University of Technology, Durban 4000, South Africa}

\author{Abdelghani Errehymy}
\email{megandhreng@dut.ac.za} \affiliation{Laboratory of High Energy Physics and Condensed Matter (LPHEMaC), Department of Physics, Faculty of Sciences A\"{i}n Chock, University of Hassan II, B.P. 5366 Maarif, Casablanca 20100, Morocco}

\begin{abstract}
In this exposition, we seek solutions of the Einstein-Maxwell field equations in the presence of a massive scalar field cast in the Brans-Dicke (BD) formalism which describes charged anisotropic strange stars. The interior spacetime is described by a spherically symmetric static metric of embedding class I. This reduces the problem to a single-generating function of the metric potential which is chosen by appealing to physics based on regularity at each interior point of the stellar interior. The resulting model is subjected to rigorous physical checks based on stability, causality and regularity. We show that our solutions describe compact objects such as PSR J1903+327; Cen X-3; EXO 1785-248 \& LMC X-4 to an excellent approximation. Novel results of our investigation reveal that the scalar field leads to higher surface charge densities which in turn affects the compactness and upper and lower values imposed by the modified Buchdahl limit for charged stars. Our results also show that the electric and scalar fields which originate from entirely different sources couple to alter physical characteristics such as mass-radius relation and surface redshift of compact objects. This superposition of the electric and scalar fields is enhanced by an increase in the BD coupling constant, $\omega_{BD}$. 
\end{abstract}

\maketitle

\section{Introduction}

Einstein's general relativity (GR) has been fruitful in describing gravitational phenomena on both cosmological and astrophysical scales. The predictions of GR has gone beyond the realms of theory and has been successfully confirmed through a plethora of experiments. With the advancement of technology these predictions have been refined. The perihelion precession of Mercury, one of the first solar system tests of GR has been drastically improved by the collection of data from Mercury
MESSENGER which orbited Mercury in 2011\cite{park}. The joint European-Japanese Mercury spacecraft BepiColombo project which launched in 2018 is expected to reveal more precise measurements of the peculiarities of Mercury's orbit. The first gravitational wave events were detected in September 2015 by the LIGO and Virgo collaborations thus reinforcing the prediction of classical GR. There is no more greater signalling of the Golden Age of astrophysical observations than the 2019 capturing of the image of the black hole at the center of galaxy M87 by the Event Horizon Telescope\cite{het}.

Despite these confirmations of GR there are still many observations that leave Einstein's classical gravity theory short. In cosmology researchers are still faced with various problems including the late-time acceleration of the Universe, dark matter and dark energy conundrums, baryon symmetry and the horizon problem, just to name a few\cite{maart1}. On the other hand there are outstanding problems in astrophysics some of which include the origin of large surface redshifts in compact objects, the behaviour of matter at extreme densities such as in the core of neutron stars and the end-states of continued gravitational collapse, amongst others. 

To this end researchers in gravitational physics have appealed to modified theories of gravitation in the hope of finding mechanisms that will account for the observations which cannot be resolved by GR. These modified theories must have as their weak field limit Einstein's general relativity. A simple modification to the standard 4D theory is to accommodate more than just the linear forms of the Riemann tensor, the Ricci tensor and the Ricci scalar in the action principle. It is well-known that incorporation of just linear tensorial quantities produces second order equations of motion which are compatible with the 4D equations\cite{love1}. The so-called Einstein-Gauss-Bonnet (EGB) gravity arises from the more general class of theories called the Lovelock polynomial Lagrangians which incorporates tensorial quantities to be of quadratic order. The beauty associated with the EGB Lagrangian is that
the equations of motion continue to be second order quasi–linear. There has been widespread interest in modeling compact objects within the framework of EGB gravity. Several exact solutions of the modified pressure isotropy condition have been derived and these models were shown to obey the stability, regularity and causality conditions required for stellar configurations. More ever, it was shown that higher order corrections alter physical properties such as compactness, stability and surface redshifts of stellar models. Recent work by Chakraborty and Dadhich on charged compact objects showed that for a given spacetime dimension, 4D stellar models are more compact than their pure Lovelock counterparts\cite{love2}. They further showed that an increase in the intensity of the electromagnetic field results in a greater compactification of the stellar object. They also demonstrated that within the context of 4D EGB gravity an increase in the strength of the Gauss-Bonnet coupling (behaving as an effective electric charge), leads to an increase in the compactness of the stellar object. 

Besides the Lovelock gravity and the popular EGB gravity theory there are a wide spectrum of other modified gravity theories which are being frequently utilised within both cosmological and astrophysical contexts. Unimodular gravity which is based on the trace-free Einstein equations was conjured to solve the magnitude of the vacuum energy density conundrum. A hugely popular modified theory of gravitation is the so-called $f(R;T)$ theory proposed by Harko et al. \cite{harko1} in which the action is the Ricci scalar
$R$ and the the trace of the energy-momentum tensor $T$.  
The Rastall theory is centered on the notion that divergence of the energy-momentum tensor is proportional to the divergence of the Ricci scalar. This has implications for the conservation of energy momentum\cite{sud1}. Starobinsky in his attempt to explain the accelerated expansion of the universe put forth a theory of gravitation whose action is quadratic in the Ricci scalar. This modified theory has come to be known as the $f(R)$ theory of gravity.  Stellar modeling in $f(R)$ gravity has received interest in the recent past which produced models which are compatible with observational data, see \cite{ellisa} and references therein. However, some recent works on the $f(R;T)$ theory, Rastall theory and Starobinsky $f(R;T)$-function can be seen in the following Refs.\cite{m11,m12,m13,m14,m15}. The Brans-Dicke (BD) gravitational theory was the first of many scalar-tensor theories of gravitation in which the non-minimally coupled scalar field represents the spacetime-varying gravitational ”constant”. The BD gravity theory successfully incorporated Mach's principle.  The BD gravity theory of gravitation is also called the Jordan–Brans–Dicke gravity theory which is a theoretical framework that can be represented in Jordan–Brans–Dicke gravity as well as Einstein's frame. It continues to be one of the more popular theories of modified classical GR and has been widely utilised in cosmological models. BD gravity has elegantly explained the inflationary epoch of the universe and the current accelerated phase of the universe without invoking any exotic matter fields or dissipative processes \cite{purba}. An emergent universe via quantum tunneling within a Jordan-Brans-Dicke framework has been recently proposed by Labrana and Cossi\cite{cos}. The initial static universe is supported by a scalar field contained within a false vacuum. The staticity of the model is broken via quantum tunneling in which the scalar field decays into a true vacuum and the universe begins to evolve dynamically. In a recent study within the Brans-Dicke framework motivated by a $f(R) = R + \alpha R^n - \delta R^{2-n}$ modified Starobinsky model inflation and a nonzero residual value for the Ricci scalar was obtained. More importantly, it was shown in the high energy limit (BD theory with a Jordan framework) predictions are consistent with data obtained by PLANCK or BICEP2\cite{mich}. On the astrophysical front recent work by several authors using the BD formalism have successfully generated models of anisotropic compact objects \cite{shar2,shar3} . Sharif and Majid\cite{shar1} obtained models of anisotropic bounded configurations via gravitational decoupling through MGD approach. They show that the stability of the model is affected by the anisotropy parameter which is inherently linked to the decoupling constant. 

We have seen a virtual explosion of exact solutions describing compact objects in classical GR and modified gravity theories. The past decade in particular has seen a proliferation of realistic stellar models which have catapulted the search for solutions of the field equations into mainstream astrophysics. Solution-generating methods have been inherently linked to physical viability tests which are backed by observational data. The Karmarkar condition has been extensively utilised to generate compact objects in which the radial and transverse stresses are unequal at each interior point of the stellar fluid\cite{jasim,ged1,rant1}. In classical GR the Karmarkar condition is immediately integrated to give a relation between the two metric potentials. This reduces the problem of finding exact solutions of the field equations to a singe-generating function\cite{k1,k11,k12,k2,k3,k4,k5}. In a recent paper, Hansraj and Moodly demonstrated that nonexistence of conformally flat charged isotropic fluid sphere of embedding class one \cite{lushen}. The embedding of the generalized Vaidya (GV) solution via the Karmarkar solution shows that embedding does not allow the interpretation of the
generalized Vaidya spacetime as a diffusive medium. In other words, the Karmarkar condition prohibits the GV solution to be interpreted as an atmosphere composed of radiation and diffusive strings of a star undergoing dissipative collapse in the form of a radial heat flux\cite{nik}. The Karmarkar condition has been extended to incorporate time-dependent systems which include modelling shear-free, dissipative collapse\cite{meg1,meg2,osp,suresh}. 
The Karmarkar condition has been successfully used in modified gravity theories to investigate contributions from the inclusion of quadratic terms of the tensorial quantities and higher dimensional effects on compactness, stability and surface redshifts of compact objects residing in these exotic spacetimes. \\
The role of an equation of state (EoS) in modeling compact objects has been highlighted in several recent studies. The simple linear EoS which expresses the pressure as a linear function of the fluid density ($p_r = \alpha\rho$) where $\alpha \geq $ is a constant has been extended to include $\alpha < 0$ used in modeling so-called dark stars and phantom fields. The MIT Bag model EoS has gained popularity amongst researchers and has been successfully utilised to model compact objects in classical GR and modified gravity theories\cite{dey1,rah1}. The quadratic EoS, polytropic EoS and Chaplygin gas EoS have also led to physically reasonable models of static stars\cite{thia,thib,bh}. By appealing to results in quantum chromodynamics and quark interactions within the stellar core, the so-called colour-flavoured-Locked (CFL) EoS has been recently used in obtaining models of compact objects which approximate realistic neutron stars, pulsars and strange stars to a very good degree\cite{thi,new12}. The CFL EoS has also been employed to study the surface tension of neutron stars. This study shows that the surface tension is sensitive to the magnitude of the Bag constant\cite{rob1}. It was observed that larger values of the Bag constant led to stellar models with lower tangential pressures and surface tensions. \\

The role of the electromagnetic field in the (in)stability of compact objects has occupied the interest of researchers since the discovery of the Schwarzschild solution. The study of charged objects in general relativity has taken a different and refreshing trajectory compared to the early attempts at just finding exact solutions to the Einstein-Maxwell system. The theoretical possibility of celestial models endowed with an electric field has been studied by several researchers \cite{bek71,zha82,fel95,fel99,maf19}. As early as 1924 Rosseland \cite{ros24} (see also Eddington \cite{edd26}), investigated the possibility that a self-gravitating stellar structure within the framework of Eddington's theory, where the stellar structure is treated as a ball of hot ionized gases containing a considerable quantity of charge. In such a system, the large number of lighter particles (electrons) as compared to positive heavier particles (ions) escape from its surface due to their higher kinetic energy and this migration of electrons continues resulting in an induced electric field within the core of the compact stellar structure. In this respect, equilibrium is achieved after a certain amount of electrons escape and the net electric charge approaches to about $100$ Coulombs per solar mass. As shown by the authors \cite{bal78}, this result applies to any bounded system whose size is smaller than the Debye length of the surrounding media. In his seminal paper addressing hydrostatic equilibrium and the gravitational collapse of charged bodies, Bekenstein argues that the densities within neutron star cores may lead to electric fields whose magnitudes may exceed the critical field required for pair creation, $10^{16}$V/cm, and hence annihilate themselves. On the other hand, the possibility of a collapsing charged stellar structure to a point singularity might be avoided by the presence of charge. Ghezzi's work on charged neutron stars showed that the outcome of gravitational collapse is sensitive to the charge to mass ratio, ${Q}/{\sqrt{G}\mu}$. Simulations based on various values of this ratio predict several possible outcomes of collapse which include the formation of black holes, naked singularities or exploding remnants. An interesting outcome of this study is that extremal black holes cannot form from the collapse of a charged fluid sphere. This huge body of work on static, charged stellar objects has provided us with more questions than answers and has prompted an invigorated interest in studying the effect of charge on the hyrdostatic equilibrium or dynamical collapse of such configurations. Herein lies the motivation for our study of charged compact objects within the BD framework with a massive scalar field. It is important to note that these toy models play an important role in checking numerical results and simulations. In this way, it is clear that stellar objects can be endowed with nonzero charge which can give rise to high intensity electric fields\cite{ma1,dassie,ma2}. The Einstein-Maxwell system can be interpreted as representing an anisotropic fluid with the pressure isotropy condition becoming the definition of the electric field. A recent paper by Maurya $\&$ Tello-Ortiz \cite{o}  highlighted an interesting interplay between the anisotropy parameter and the electric field intensity which provides a mechanism for maintaining stability of the stellar configuration. They found that the force due to the pressure anisotropy initially dominates the
Coulombic repulsion closer to the center of the star with the anisotropic force out-growing the Coulombic repulsion towards the surface layers of the star. A similar phenomenon was discovered in a charged compact star of embedding class 1\cite{mago}. The effect of charge on stellar characteristics within the framework of Einstein-Gauss-Bonnet (EGB) gravity has been investigated. It is found that mass-radius relation and surface redshifts are modified by the presence of the EGB coupling constant\cite{bharg}.

This paper is structured as follows:
In \S2. we provide the necessary equations within the BD formalism necessary to model a charged compact object. The class 1 embedding condition is derived for the BD framework in \S3. The junction conditions required for the smooth matching of the interior spacetime to the Reissner-Nordstrom exterior is presented in \S4. In \S5 we discuss the regularity of the metric functions and thermodynamical quantities at the center of the stellar configuration  and we derive the modified TOV equation in the presence of a massive scalar field, nonzero charge density and pressure anisotropy together with mass-radius relation and moment of inertia thorough $M-R$ and $M-I$ curves in \S5. A detailed discussion of the physical attributes together with the conclusion of our model follows in \S6.

\section{The background of Brans-Dicke gravity theory and Field equations}
  The action of scalar–tensor theory in Brans–Dicke frame in  relativistic units $G=c=1$ is defined as,
\begin{eqnarray}\label{eq1}
\hspace{-0.5cm} S= \frac{1}{16\pi}\int \bigg[\mathcal{R}\Phi-\frac{\omega_{BD}}{\Phi} \nabla^i\nabla_i \Phi-\mathcal{L}(\Phi)\bigg]\sqrt{-g}~ d^{4}x+\int \mathcal{L}_m \sqrt{-g}~ d^{4}x+\int \mathcal{L}_e \sqrt{-g}~ d^{4}x, 
\end{eqnarray}
where $\mathcal{R}$, $g$, $\mathcal{L}_m$, and $\mathcal{L}_e$ describe the Ricci scalar, determinant of metric tensor, matter Lagrangian density, and Lagrangian electromagnetic field respectively, while $\omega_{BD}$ is a dimensionless Dicke coupling constant and $\Phi$ is a scalar field. Here the function $\mathcal{L}(\Phi)$ depends completely on  the scalar field $\Phi$. In the present case we define this scalar field function $\mathcal{L}(\Phi)$ as,
\begin{eqnarray}\label{eq2}
\mathcal{L}(\Phi)=\frac{1}{2} m_{\phi}^2 \Phi^2
\end{eqnarray}
Now by varying of the action (\ref{eq1}) with respect to the metric tensor $g^{ij}$ and scalar field $\Phi$ provides the following field equations and evaluation equation, respectively, which can be written as,
\begin{eqnarray}
G_{ij}=\frac{1}{\Phi}\bigg[8\pi T^{(m)}_{ij}+8\pi E_{ij}+T^{(\Phi)}_{ij}\bigg],\label{eq3}
\end{eqnarray}
where, $T^{(m)}_{ij}$ and $E_{ij}$ denote the energy-momentum tensor for matter distribution and electromagnetic field tensor, respectively while $T^{(\Phi)}_{ij}$ represents a scalar tensor appearing in the system due to the scalar field $\Phi$. All the field tensors can be written as,
\begin{eqnarray}
&& T^{(m)}_{ij}=(\rho+p_t) u_i u_j-p_t g_{ij}+(p_r-p_t)v_iv_j,\label{eq4}\\
&& E_{ij}=\frac{1}{4 \pi} \left(-F^{n}_{i}F_{j\,n} + \frac{1}{4}{g_{ij}} F_{\gamma\,n}F^{\gamma\,n} \right), \label{eq4a}\\
&& T^{(\Phi)}_{ij}=\Phi_{,i;j}-g_{ij} \Box \Phi +\frac{\omega_{BD}}{\Phi} \bigg(\Phi_{,i}\, \Phi_{,j} -\frac{g_{ij} \Phi_{,\delta}\,\Phi^{,\delta}}{2}\bigg) -\frac{\mathcal{L}(\Phi)\,g_{ij}}{2}. \label{eq5}
\end{eqnarray}
Here, $\Box$ denotes the d'Alembert operator, then $\Box \Phi$ can be given as,
\begin{eqnarray}
\Box \Phi= \frac{T^{(m)}}{3+2\,\omega_{BD}}+\frac{1}{3+2\, \omega_{BD}} \bigg(\Phi \frac{d \mathcal{L}(\Phi)}{d\Phi}-2 \mathcal{L}(\Phi) \bigg), \label{eq6}
\end{eqnarray}
Here, $\rho$, $p_r$ and $p_t$ denote the energy density, radial pressure and transverse pressure, respectively with $T^{(m)}$ being the trace of the energy tensor $T^{(m)}_{ij}$.In addition, the anti-symmetric electromagnetic field tensor $F_{i j}$ given in Eq. (\ref{eq4a}) is characterized as 
\begin{eqnarray}
F_{ij}=\nabla_{i}\,A_{j}-\nabla_{j}\,A_{i}
\end{eqnarray}
for which Maxwell's equations have been satisfied,
\begin{eqnarray}
F_{i j,k} + F_{j k,i}+F_{k i, j}=0
\end{eqnarray}
with
\begin{equation}
 F^{ik}\,_{;k} = 4\pi\,J^i  \label{8a}
\end{equation}
where, $J^i$ is the electromagnetic 4-current vector. This can be expressed as
\begin{eqnarray}
J^{i}=\frac{\sigma}{\sqrt{g_{00}}}\,\frac{dx^i}{dx^0}=\sigma\,u^i, \label{8b}
\end{eqnarray}
where $\sigma=e^{\xi/2}\,J^{0}(r)$ representing the charge density. It turns out that for a static matter distribution with spherical symmetry, there is only one non-zero component of the electromagnetic 4-current $J^i$ which is $J^0$, a function of the radial distance, $r$. The $F^{01}$ and $F^{10}$ components are the only non-zero components of electromagnetic field tensor expressed in (\ref{eq4a}) and they are connected by the formula $F^{01}=-F^{10}$, which characterizes the radial constituent of the electric field. The constituent of the electric field is determined through Eqs. (\ref{8a}) and (\ref{8b}) as follows
\begin{eqnarray}
F^{01} = - F^{10}=\frac{q}{r^2}\,e^{-(\xi+\eta)/2}
\end{eqnarray}
The quantity $q(r)$ represents the effective charge of a spherical system of radial coordinate, $r$, subsequently, this electric charge can be characterized by the relativistic Gauss law and corresponding electric field $E$ explicitly as,
\begin{eqnarray}
q(r)&=&4\,\pi\int^{r}_{0}{\sigma\,r^2\,e^{\eta/2} dr}=r^2\,\sqrt{-F_{10}\,F^{10}}\\
E^2&=&-F_{10}\,F^{10}=\frac{q^2}{r^4}.
\end{eqnarray}
It is noted that Doneva et al. \cite{Doneva} and Yazadjiev et al. \cite{Yazadjiev} have already discussed both slowly and rapidly rotating neutron stars by using the above potential function (\ref{eq2}). 
In order to describe the stellar structure, we assume a static spherically symmetric line element which can be cast as,
\begin{eqnarray}
ds^2=e^{\xi(r) } \, dt^{2}-e^{\eta(r)} dr^{2}-r^{2}(d\theta ^{2} +\sin ^{2} \theta \, d\phi ^{2}), \label{eq7}
\end{eqnarray}
where $\xi(r)$ and $\eta(r)$ are metric potentials and rely just upon the radial distance $r$ that ensures the staticity of the space-time. Also, $u^i=e^{-\xi/2}\delta^i_0$, designating the four-velocity, and $v^i=e^{-\eta/2}\delta^i_1$, designating the unit space like vector, which are specified as, $u^iu_i=1$ and $v^iv_i=-1$, and $\Box \Phi=\Phi^{,i}_{; i}=(-g)^{-\frac{1}{2}} [(-g)^{-\frac{1}{2}}\,\Phi^{,i}]_{,i}$. By using Eqs. (\ref{eq2}) - (\ref{eq7}), we obtain the following field equations,

\begin{eqnarray}
&& {{\rm e}^{-\eta }}\left( {\frac {\eta^{{\prime}}}{r}}-\frac{1}{r^2}\right)+\frac{1}{r^2}=\frac{1}{\Phi} \bigg(8\pi \rho+\frac{q^2}{r^4}+T^{0(\Phi)}_0 \bigg),\label{eq8}\\
&& {{\rm e}^{-\eta}} \left( {\frac {\xi^{{\prime}}}{r}}+\frac{1}{r^2}\right) -\frac{1}{r^2}=\frac{1}{\Phi} \bigg( 8\pi p_r-\frac{q^2}{r^4}-T^{1(\Phi)}_1 \bigg),\label{eq9}\\
&& \frac{{\rm e}^{-\eta}}{2} \left( \xi^{{\prime\prime}}+\frac{{\xi^{{\prime}}}^{2}}{2}+{\frac {\xi^{{\prime}}-\eta^{{\prime}}}{r}}-\frac{\xi^{{\prime}}\eta^{{\prime}}}{2} \right) =\frac{1}{\Phi} \bigg(8\pi p_t+\frac{q^2}{r^4}-T^{2(\Phi)}_2 \bigg),\label{eq10}
\end{eqnarray}
where prime denotes differentiation with respect to $r$. On the other hand, the scalar tensor components $T^{0(\Phi)}_0$, $T^{1(\Phi)}_1$, and $T^{2(\Phi)}_2$ in terms of $\xi$ and $\eta$ are given as,

\begin{eqnarray}
&& T^{0(\Phi)}_0= e^{-\eta} \bigg[\Phi^{\prime\prime}+\bigg(\frac{2}{r}-\frac{\eta^{\prime}}{2}\bigg)\Phi^{\prime}+\frac{\omega_{BD}}{2\Phi} \Phi^{\prime 2}-e^{\eta}\,\frac{\mathcal{L}(\Phi)}{2} \bigg],\\
&& T^{1(\Phi)}_1= e^{-\eta} \bigg[\bigg(\frac{2}{r}+\frac{\xi^{\prime}}{2}\bigg)\Phi^{\prime}-\frac{\omega_{BD}}{2\Phi} \Phi^{\prime 2}-e^{\eta}\,\frac{\mathcal{L}(\Phi)}{2} \bigg],\\
&& T^{2(\Phi)}_2= e^{-\eta} \bigg[\Phi^{\prime\prime}+\bigg(\frac{1}{r}-\frac{\eta^{\prime}}{2}+\frac{\xi^{\prime}}{2}\bigg)\Phi^{\prime}+\frac{\omega_{BD}}{2\Phi} \Phi^{\prime 2}-e^{\eta}\,\frac{\mathcal{L}(\Phi)}{2} \bigg]. 
\end{eqnarray}
However, from Eqs.(\ref{eq6}) and (\ref{eq7}) we obtain, \begin{eqnarray}
\hspace{-0.5cm}\Box \Phi &=& -e^{-\eta} \bigg[\bigg(\frac{2}{r}-\frac{\eta^{\prime}}{2}+\frac{\xi^{\prime}}{2}\bigg)\Phi^{\prime}(r)+\Phi^{\prime\prime}(r)\bigg]=\frac{1}{(3+2\,\omega_{BD})}\bigg[T^{(m)}+\Phi \frac{d \mathcal{L}(\Phi)}{d\Phi}-2\,\mathcal{L}(\Phi)\bigg].~~~ 
\end{eqnarray}
It has been argued that extreme temperatures and pressures at the core of massive neutron stars can transform into quark stars with up ($u$), down ($d$) and strange star ($s$) quark flavors. In this regard, we suppose that the MIT bag model rules the matter variables (density and pressure) in the interior of these relativistic massive stars. In addition, we assume that non-interacting and massless quarks occupy the inside of the stellar structures. According to the MIT Bag model the quark pressure $p_r$ can be cast as 
\begin{eqnarray} \label{eq16}
p_r=\sum_{f} p^{f}-\mathcal{B},~~~~f=u,~d,~s
\end{eqnarray}
where $p^{f}$ describes the individual pressures due to each quark flavor which is balanced by the Bag constant (or total external Bag pressure) $\mathcal{B}$. The deconfined quarks inside the MIT Bag model have the accompanying total energy density
\begin{eqnarray}
\rho=\sum_{f} \rho^{f}+\mathcal{B},\label{eq17}
\end{eqnarray}
where $\rho^f$ indicates the matter density due to each flavor which is connected to the corresponding pressure by the formula given as $\rho^f=3\,p^f$. Consequently, Eqs. (\ref{eq16}) and (\ref{eq17}) are consolidated to express the following simplified MIT Bag model,
\begin{eqnarray} \label{18}
p_r=\frac{1}{3}(\rho-4\mathcal{B}),  
\end{eqnarray}
It should be mentioned here that this specific linear form of the MIT bag model EoS has been applied for portraying the stellar systems made of the strange quark matter distribution in pure  GR and modified gravity theories. 

Now using the Eqs. (\ref{eq8}) and (\ref{eq9}) along with EOS (\ref{18}), we obtain the expression for the electric field as,
\begin{eqnarray}
\frac{4q^2}{r^4}=\Phi\,\bigg[{{\rm e}^{-\eta }}\left( {\frac {\eta^{{\prime}}-3\xi^{{\prime}}}{r}}\right)+\frac{4\,(1-e^{-\eta})}{r^2}\bigg]-\big(3T^{1(\Phi)}_1+T^{0(\Phi)}_0\big)-32 \pi \mathcal{B},  \label{e2.27}
\end{eqnarray}

\section{Basic formulation of Class one condition and its solution in Brans-Dicke gravity }
It is well-known that the embedding of $n-$dimensional space $V^{n}$ in a pseudo-Euclidean space $E^{n}$ attracted much consideration as inferred by Eisland \cite{eisland} and Eisenhart \cite{eisenhart}. In the case where a $n-$dimensional  space $V^{n}$ can be isometrically immersed in $(n+m)-$dimensional space, where $m$ is a minimum number of supplementary dimensions, at this stage $V^{n}$ is said to be $m-$class embedding. Habitually, the metric expressed in (\ref{eq7}) provides the four$-$dimensional spherically symmetric space-time which describes a space-time of class two i.e, when $m=2$, which shows that it is embedded in a six$-$dimensional pseudo-Euclidean space. On the other hand, it should be noted that one can reveal a possible parametrization in order to incorporate the space-time expressed in (\ref{eq7}) into a five$-$dimensional pseudo-Euclidean space which leads to class $m=1$ known as embedding class one~\cite{eisland,eisenhart,karmarkar}. For a spherically symmetric space-time in both cases static or non-static to be class-one, the system has to be consistent with the following necessary and suitable conditions:
\begin{itemize}
 \item For a stellar system, the symmetric tensor $b_{ij}$ should be determined under the associated Gauss conditions:
\begin{eqnarray} \label{eq16a}
\mathcal{R}_{ijhk}=\epsilon\Big(b_{ih}b_{jk}-b_{ik}b_{jh}\Big),   \end{eqnarray}
where $\epsilon=\pm1$ is everywhere normal to the manifold is time-like (+1) or space-like (-1).
\item The symmetric tensor $b_{ij}$ must fulfill the following differential equation, known as Codazzi equation, as:
\begin{equation}\label{eq17a}
\nabla_{h}b_{ij}-\nabla_{i}b_{hj}=0.
\end{equation}
\end{itemize}
Now the Riemann components for the line element (\ref{eq7}) can be expressed as follows, 
\begin{eqnarray}
&&\hspace{-1cm} \mathcal{R}_{0101}=-e^{\xi}\left(\frac{\xi^{\prime\prime}}{2}-\frac{\eta^{\prime}\xi^{\prime}}{4}+\frac{\xi^{\prime2}}{4}\right); ~~ \mathcal{R}_{1313}=-\frac{r}{2}\eta^{\prime}\sin^2\theta; ~~ \mathcal{R}_{2323}=-\frac{r^{2}\sin^{2}\theta}{e^{\eta}}\Big(e^{\eta}-1\Big); \nonumber\\
&&\hspace{-1cm}\mathcal{R}_{1202}=0,~~~
  \mathcal{R}_{0202}=-\frac{r}{2}\xi^{\prime}e^{\xi-\eta};~~ 
\mathcal{R}_{1303}=0;~~\mathcal{R}_{0303}=-\frac{r}{2}\xi^{\prime}e^{\xi-\eta} \sin^2\theta;~~ \mathcal{R}_{1212}=-\frac{r}{2}\eta^{\prime}.
\end{eqnarray}
Substituting these Riemann components into Gauss's equation (\ref{eq16a}) leads to 
\begin{eqnarray}\label{Eq3.4}
&&\hspace{-1.cm}
b_{01}b_{33}=\mathcal{R}_{1303}=0;~~
b_{01}b_{22}=\mathcal{R}_{1212}=0;
b_{00}b_{33}=\mathcal{R}_{0303};
b_{00}b_{22}=\mathcal{R}_{0202}; \nonumber\\
&&\hspace{-1cm}
b_{11}b_{33}=\mathcal{R}_{1313}; ~~~
b_{22}b_{33}=\mathcal{R}_{2323}; ~~
b_{11}b_{22}=\mathcal{R}_{1212}; ~~
b_{00}b_{11}=\mathcal{R}_{0101}.
\end{eqnarray}
These relations expressed in (\ref{Eq3.4}) lead immediately to the following expressions
\begin{eqnarray}
\label{eq18}
&&\hspace{-0.7cm} \left(b_{00}\right)^{2}=\frac{\left(\mathcal{R}_{0202}\right)^{2}}{\mathcal{R}_{2323}}\sin^{2}\theta;~\left(b_{11}\right)^{2}=\frac{\left(\mathcal{R}_{1212}\right)^{2}}{\mathcal{R}_{2323}}\sin^{2}\theta;~\left(b_{22}\right)^{2}=\frac{\mathcal{R}_{2323}}{\sin^{2}\theta}; ~ \left(b_{33}\right)^{2}=\sin^{2}\theta~ \mathcal{R}_{2323}.~~~~~
\end{eqnarray} 
By combining the last term of the relation (\ref{Eq3.4}) together with the components of Eq. (\ref{eq18}), we found the following relationship in terms of the Riemann components
\begin{equation}\label{eq19}
\mathcal{R}_{0202}\mathcal{R}_{1313}=\mathcal{R}_{0101}\mathcal{R}_{2323},
\end{equation}
subject to $\mathcal{R}_{2323}\neq0$~(Pandey and Sharma condition \cite{sharma}). It ought to be noticed that all the components are given in (\ref{eq18}) fulfill the Codazzi equation (\ref{eq17a}). On the other hand, in the case of a general non--static spherically symmetric space--time, the relation between components for symmetric tensor $b_{ij}$ and Riemann tensor $R_{ijhk}$ can be given as follows
\begin{equation}\label{eq20}
b_{01}b_{22}=\mathcal{R}_{1212} \quad \mbox{and}\quad  b_{00}b_{11}-\left(b_{01}\right)^{2}=\mathcal{R}_{0101},  
\end{equation}
where $\left(b_{01}\right)^{2}=\sin^{2}\theta \left(\mathcal{R}_{1202}\right)^{2}/\mathcal{R}_{2323}$. In this situation, the embedding Class-one condition known as Karmarkar condition \cite{eisland,karmarkar} take the following form
\begin{equation}\label{eq21}
\mathcal{R}_{0202}\mathcal{R}_{1313}=\mathcal{R}_{0101}\mathcal{R}_{2323}+\mathcal{R}_{1202}\mathcal{R}_{1303}.    
\end{equation}
Although in our circumstance, the relation (\ref{eq19}) between Riemann components under the static spherically symmetric line element (\ref{eq7}) will be equivalent to relation (\ref{eq21}). The condition (\ref{eq21}) plays a fundamental role for describing the space-time (\ref{eq7}) to be class-one, which is also well-known as a necessary and sufficient condition. At this point, by incorporating the Riemann components in expression (\ref{eq21}), we obtain the following differential equation,
\begin{equation}\label{eq22}
\frac{2\xi''}{\xi^{\prime}}+\xi'=\frac{\eta'e^{\eta}}{e^{\eta}-1},
\end{equation}
with $e^{\eta}\neq 1$. By integrating equation (\ref{eq21}), we find the relationship amongst the gravitational potentials in the following form
\begin{equation}
\xi(r)=2 \ln\left[A+B\int\sqrt{e^{\eta}-1}\,dr\right].\label{e3.10}
\end{equation}
Here $A$ and $B$ are integration constants. Moreover, the solution determined by the relation (\ref{e3.10}) is called a embedding class one solution for the line element (\ref{eq7}). It is mentioned that the above method has been utilized to model compact stellar objects in various contexts.  
In order to find the solution of the field Eqs. (\ref{eq8})-(\ref{eq10}) in Brans-Dicke gravity under Class I condition (\ref{e3.10}), we need to find the metric potentials admitting Karmarkar condition with Pandey-Sharma~  condition \cite{sharma}  ($\mathcal{R}_{2323}\neq0$). As it is well-known in general, the invariance of the Ricci tensor necessitates that the matter variables viz., energy density $\rho$, radial pressure $p_r$ and transverse pressure $p_t$ ought to be finite at the center. The regularity of the Weyl invariant  necessitates that the following two quantities: mass $m(r)$ and electric charge $q(r)$ must satisfy the following conditions: $m(0)=q(0)=0$ and $m(0)=0$, $m^{\prime}(r)> 0$ and $q(0)=0$, $q^{\prime}(r) > 0$ i.e., both said quantities reach their minimum and maximum values at the center as well as at the surface of the celestial body, respectively. In this regard, Maurya and collaborators \cite{maurya2017} have already demonstrated that the gravitational potential $\xi(0)$ is equal to a finite constant value, $q(0) = 0$, $\xi^{\prime}(0) = 0$ and $\xi^{\prime\prime}(0)>0$ according to the modelling of charged anisotropic compact celestial bodies. Since both physical quantities viz., energy density $\rho$ and radial pressure $p_r$ are positive finite and continuous and also pursue the condition $r>2m(r)$ \cite{r31,r32}. From $p_r(r)\ge 0$ with the help of the condition $r>2m(r)$, we can obtain $\xi^\prime\ne0$, which implies that the general function $\xi(r)$ is regular minimum at the center and a monotonic increasing function of the radial coordinate $r$. Consequently, the general function $\xi(r)$ should conserve the said physical characteristics. Moreover, the gravitational potential function $e^{\eta}$ should fulfill the accompanying form $e^{\eta}=1+O(r^2)$ to ensure the regularity and stability of the compact stellar object. In this way by keeping all the attributes in our mind,
we have assumed the $e^\eta$ as follow:
\begin{equation}
e^{\eta(r)} = 1+a r^2 \, e^{b \,r^2}. \label{e4.1}
\end{equation}
Using \eqref{e4.1} in \eqref{e3.10}, we get
\begin{eqnarray}
\xi(r) = 2 \ln \left[A+C \, e^{\,b r^2/2}\right].
\end{eqnarray}
{where, $C=\frac{\sqrt{a}\, B}{b}$. Furthermore, the explicit expression for energy density and pressure can be given as,
\begin{eqnarray}
8\pi\rho&=&\frac{1}{4\Phi\, (A + C e^{b r^2/2})\, (1 + a e^{b r^2} r^2)^2}\bigg(C\,e^{b r^2/2}\big[6 a e^{b r^2} \Phi^2 + 32 \mathcal{B} \Phi \pi + 3 a \Phi^{\prime} e^{b r^2} \Phi r + \big(2+ar^2e^{b r^2}\big)\nonumber\\&& a \mathcal{B} e^{b r^2} \Phi \pi r^2 -3 \Phi^{\prime\prime} \Phi (1 + a e^{b r^2} r^2)+ 3 b \Phi (2 \Phi + \Phi^{\prime} r)(1 + 2 a e^{b r^2} r^2) - 3 \Phi^{\prime2} \omega_{BD} - 3 a \Phi^{\prime2} e^{b r^2} \nonumber\\&& r^2 \omega_{BD}\big]+ A\,  [32 \mathcal{B} \Phi \pi + 32 a^2 \mathcal{B} e^{2b r^2} \Phi \pi r^4 - 3 \Phi^{\prime\prime} \Phi (1 + a e^{b r^2} r^2)-3 \Phi^{\prime2} \omega_{BD} + 
 a e^{b r^2} \{6 \Phi^2 \nonumber\\&& (1 + b r^2) + 
    \Phi r (3 \Phi^{\prime}  + 64 \mathcal{B} \pi r + 3 b \Phi^{\prime} r^2) - 3 \Phi^{\prime2} r^2 \omega_{BD}\}]\bigg),\\
 8\pi p_r&=&\frac{8\,\pi}{3} ( \rho-4\,\mathcal{B}),\\
 8\pi p_t &=&\frac{1}{4\Phi\,r\, (A + C e^{b r^2/2})\, (1 + a e^{b r^2} r^2)^2}\bigg(C\,e^{b r^2/2} \big[\Phi^{\prime} \Phi \,(12 + 7\, b\,r^2 + 7\,a\,e^{b r^2}\, r^2 + 2\,a\,b\,e^{b r^2} r^4) \nonumber\\&& +  \Phi\, r \,\big(5 \Phi^{\prime\prime} + 14\,b\,\Phi - 10\,a\,e^{b r^2} \Phi - 2 m^2_{\phi} \Phi^2 + 32 \mathcal{B} \pi + 5 a \Phi^{\prime\prime} e^{b r^2} r^2 + 4 b^2 \Phi r^2 + 4\, a\, b\, e^{b r^2} \Phi\, r^2 \nonumber\\&& - 4 \,a^2\, e^{2b r^2} \Phi\, r^2 - 4\, a\, e^{b r^2} m^2_{\phi} \Phi^2\, r^2 + 64\, a \, \mathcal{B}\, e^{b r^2} \pi\, r^2 - 2 a^2 \, e^{2b r^2}\, m^2_{\phi} \Phi^2\, r^4 + 32 \,a^2\, \mathcal{B}\, e^{2b r^2}\,\pi\, r^4\big) \nonumber\\&&+ 
   \Phi^{\prime2} r (1 + a e^{b r^2} r^2) \omega_{BD} \big] +  A \big\{-\Phi^{\prime} \Phi [-12 + a e^{b r^2} r^2 (-7 + 5 b r^2)] + \Phi\, r\,\big\{5 \Phi^{\prime\prime} (1 + a e^{b r^2} r^2) \nonumber\\&&- 2 (m^2_{\phi}\Phi^2 - 16 \mathcal{B} \pi + a e^{b r^2} (5 \Phi + 3 b \Phi r^2 + 2 m^2_{\phi} \Phi^2 r^2 - 32 \mathcal{B} \pi r^2) + a^2 e^{2b r^2} r^2 (2 \Phi + m^2_{\phi} \Phi^2 r^2 \nonumber\\&& - 16 \mathcal{B} \pi r^2))\big] + \Phi^{\prime2} r (1 + a\, e^{b r^2} r^2) \omega_{BD}\big\}\bigg).
\end{eqnarray}
Consequently, the expression for the electric field intensity can be obatined from the Eq.(\ref{e2.27}) as,
\begin{eqnarray}
E^2&=&\frac{1}{4\Phi\,r\, (A + C e^{b r^2/2})\, (1 + a e^{b r^2} r^2)^2}\bigg(C\,e^{b r^2/2} \big[-\Phi\, r\, (\Phi^{\prime\prime} + 6 b \Phi - 6 a e^{b r^2} \Phi - m^2_{\phi} \Phi^2 + 
    32\, \mathcal{B}\, \pi \nonumber\\&&+ a \Phi^{\prime\prime} e^{b r^2} r^2 + 4 a b e^{b r^2} \Phi r^2 - 
    4 a^2 e^{2b r^2} \Phi r^2 - 2 a e^{b r^2} m^2_{\phi} \Phi^2 r^2 + 
    64 a \mathcal{B} e^{b r^2} \pi\, r^2 - a^2\, e^{2b r^2} m^2_{\phi}\, \Phi^2\, r^4  \nonumber\\&&+ 
    32 a^2 \mathcal{B} e^{2b r^2} \pi r^4) - 
 \Phi^{\prime} \Phi (8 + 7 a e^{b r^2} r^2 + b r^2 (3 + 2 a e^{b r^2} r^2)) + 
 \Phi^{\prime2} r (1 + a e^{b r^2} r^2) \omega_{BD}\big]  +  A \big\{\Phi^{\prime}\, \Phi\nonumber\\&& (-8 + a e^{b r^2} r^2 (-7 + b r^2)) + \Phi\, r\, \big[m^2_{\phi} \Phi^2 - 32 \mathcal{B} \pi - \Phi^{\prime\prime} (1 + a e^{b r^2} r^2) + a^2 e^{2b r^2}\, r^2\, (4 \Phi  + m^2_{\phi} \Phi^2 r^2\nonumber\\&& - 32 \mathcal{B} \pi r^2) + 
    2 a e^{b r^2} (3 \Phi + b \Phi r^2 + m^2_{\phi} \Phi^2 r^2 - 32 Bg \pi r^2)\big] + \Phi^{\prime2} r (1 + a e^{2b r^2} r^2) \omega_{BD}\big\}\bigg).
\end{eqnarray}}
\section{Junction conditions}
To describe the complete structure of the self gravitating anisotropic compact object, the interior spacetime must be matched smoothly with the exterior spacetime at the 
pressure-free boundary $\Sigma$. The exterior spacetime is considered to be the Reissner-Nordstrom spacetime given by,
\begin{eqnarray}
ds^2_+= - \bigg(1-\frac{2{\mathcal{M}}}{r}+\frac{Q^2}{r^2}\bigg)^{-1} dr^2-r^2(d\theta^2-\sin^2\theta\,d\phi^2)+\bigg(1-\frac{2{\mathcal{M}}}{r}+\frac{Q^2}{r^2}\bigg)\,dt^2,~\label{metric2}
\end{eqnarray} 
where, $M$ is the total mass. In order to satisfy the smoothness and continuity of internal spacetime metric $ds^2_-$ and the external spacetime $ds^2_+$ at the boundary surface $\Sigma$, the following conditions must be fulfilled at the hypersurface $\Sigma$ ($f=r-R=0$, {where}~$R$ is a radius). 
\begin{eqnarray}
&& [{ds^2}_{-}]_{\Sigma} = [{ds^2}_{+}]_{\Sigma},~~~~~~~[{K_{ij}}_{-}]_{\Sigma}=[{K_{ij}}_{+}]_{\Sigma}, \label{boun1} \\
&& [\Phi(r)_{-}]_{\Sigma} = [\Phi(r)_{+}]_{\Sigma},~~~~~~[\Phi^{\prime}(r)_{-}]_{\Sigma}=[\Phi^{\prime}(r)_{+}]_{\Sigma}. \label{boun2}
\end{eqnarray}
Here, as usual $-$ and $+$ denote the interior and exterior spacetimes, respectively while $K_{ij}$ represents the curvature. 
By using the continuity of the first fundamental form ($[ds^2]_{\Sigma}$=0), we will always get
\begin{eqnarray}
[F]_{\Sigma} \equiv F (r\longrightarrow R^{+})-F (r\longrightarrow R^{-})\equiv F^{+}(R)-F^{-}(R)
\end{eqnarray}
for any function $F(r)$. This condition provides us $g^{-}_{rr}(R)=g^{+}_{rr}(R)$ and $g^{-}_{tt}(R)=g^{+}_{tt}(R)$. On the other hand, the spacetime (\ref{eq7}) must satisfy the second fundamental form ($K_{ij}$) at the surface $\Sigma$ which is equivalent to the O$'$Brien and Synge \cite{Brien} junction condition. This condition says that radial pressure must be zero at boundary $r=r_{\Sigma}$, which leads
\begin{eqnarray} \label{boun3}
p_r(R)=0. \label{eq5.5}
\end{eqnarray}
{In addition to above, the BD scalar field $\Phi$ corresponding to the vacuum Schwarzschild solution is derived using the technique in [46]. We consider $\tau^{-}$ and $\tau^{+}$ as the inner and outer space, respectively.
Then the hypersurface is defined by the following line element 
\begin{eqnarray}
ds^2=d\gamma^2-R^2(d\theta^2+\sin^2\theta\,d\phi^2). \label{eq4.6}
\end{eqnarray}
where $\gamma$ defines the proper time boundary. The extrinsic curvature of this boundary $\Sigma$ can be written as,
\begin{eqnarray}
K^{\pm}_{ij}=-m_k^{\pm}\,\frac{\partial^2y^k_{\pm}}{\partial n^i\,n^j}-m_k^{\pm}\,\Gamma^k_{\mu\,l}\frac{\partial y^\mu_{\pm}}{\partial n^i}\frac{\partial y^l_{\pm}}{\partial n^j}
\end{eqnarray}
Here the symbol $n^i$ defines the coordinates in the boundary $\Sigma$, while $m_k^{\pm}$ denotes the four-velocity normal to $\Sigma$. The components of this four-velocity is given in the coordinates ($ y^\nu_{\pm}$) of $\tau^{\pm}$ as  
\begin{eqnarray}
m_k^{\pm}=\pm\,\frac{df}{dy^{k}}\,\big|g^{\mu\,l}\frac{df}{dy^{\mu}}\frac{df}{dy^{l}}\big|^{-\frac{1}{2}} ~~~~~\text{with}~~~m_km^k=1
\end{eqnarray}
Then the unit normal vectors for inner and outer regions can be cast as
\begin{eqnarray}
m^{-}_k=\bigg[0,e^{\eta/2},0,0\bigg],~~~\text{and}~~~~m^{-}_k=\bigg[0,\bigg(1-\frac{2\mathcal{M}}{r}+\frac{Q^2}{r^2}\bigg)^{-1/2},0,0\bigg].\label{eq4.9}
\end{eqnarray}
In view of line elements (\ref{eq7}) and (\ref{eq4.6}) together with the Reisner-Nordstrom spacetime (\ref{metric2}), we can write
\begin{eqnarray}
\bigg[\frac{dt}{d\gamma}\bigg]_{\Sigma}=\big[e^{-\frac{\xi}{2}}\big]_{\Sigma}=\bigg[\bigg(1-\frac{2\mathcal{M}}{r}+\frac{Q^2}{r^2}\bigg)^{-\frac{1}{2}}\bigg]_{\Sigma},~~~\text{where}~~~~[r]_{\Sigma}=R. \label{eq4.10}
\end{eqnarray}
The non-zero components of the curvature $(K_{ij})$ can be obtained from Eq.(\ref{eq4.9}) as,
\begin{eqnarray}
&& K^{-}_{00}=\bigg[-\frac{\xi^{\prime}}{2\,e^{\eta/2}}\,\bigg]_{\Sigma},~~~~~~~ K^{-}_{22}=\frac{1}{\sin^2\theta}\,K^{-}_{33}= \big[r\,e^{-\eta/2}\big]_{\Sigma},\nonumber\\
&& K^{+}_{00}=\bigg[\bigg(\frac{\mathcal{M}}{r^2}-\frac{Q^2}{r^3}\bigg)\bigg(1-\frac{2\mathcal{M}}{r}+\frac{Q^2}{r^2}\bigg)^{-\frac{1}{2}}\bigg]_{\Sigma},~~K^{+}_{22}=\frac{1}{\sin^2\theta}\,K^{+}_{33}=\bigg[r\,\bigg(1-\frac{2\mathcal{M}}{r}+\frac{Q^2}{r^2}\bigg)^{\frac{1}{2}}\bigg]_{\Sigma}.~~~~~\nonumber
\end{eqnarray}
Then the junction condition $[K^{-}_{22}]_{\Sigma}=[K^{+}_{22}]_{\Sigma}$ together with $[r]_{\Sigma}$ provides,
\begin{eqnarray}
e^{-\eta(R)}=\bigg(1-\frac{2\mathcal{M}}{R}+\frac{Q^2}{R^2}\bigg). \label{eq4.11}
\end{eqnarray}
Plugging the above expression into the matching condition $[K^{-}_{00}]_{\Sigma}=[K^{+}_{00}]_{\Sigma}$ yields,
\begin{eqnarray}
\xi^\prime(R)=\bigg(\frac{\mathcal{M}}{R^2}-\frac{Q^2}{R^3}\bigg)\bigg(1-\frac{2\mathcal{M}}{R}+\frac{Q^2}{R^2}\bigg)^{-1}. \label{eq4.12}
\end{eqnarray}
Therefore, the matching conditions given by Eqs.(\ref{eq4.10})-(\ref{eq4.12}) leads to the following relations at the hypersurface,
\begin{eqnarray}
 \big[A+C\,e^{bR^2/2}\big]&=&\bigg(1-\frac{2\mathcal{M}}{R}+\frac{Q^2}{R^2}\bigg)^{1/2},\label{eq4.13}\\
 {1+aR^2\,e^{bR^2}}&=&\bigg(1-\frac{2\mathcal{M}}{R}+\frac{Q^2}{R^2}\bigg)^{-1},\label{eq4.14}\\
 2C\,bR\,e^{bR^2/2}&=&\bigg(\frac{\mathcal{M}}{R^2}-\frac{Q^2}{R^3}\bigg)\bigg(1-\frac{2\mathcal{M}}{R}+\frac{Q^2}{R^2}\bigg)^{-1/2}.\label{eq4.15}
\end{eqnarray}
After solving the relations (\ref{eq4.13})-(\ref{eq4.15}), we obtain expressions for the constants,
\begin{eqnarray}
a &=& \frac{e^{-b R^2} \left(2 \mathcal{M} R-Q^2\right)}{R^2 [R (R-2 \mathcal{M})+Q^2]},\\
A &=& \sqrt{\frac{R^2-2 \mathcal{M} R+Q^2}{R^2}}-\frac{\sqrt{a} B e^{b R^2/2}}{b},\\
B &=& \frac{e^{-b R^2/2} \left(\mathcal{M} R-Q^2\right)}{2 \sqrt{a}\, R^4} \sqrt{\frac{R^2}{R^2-2 \mathcal{M} R+Q^2}}.
\end{eqnarray}
}

\section{Physical viability of the model}
We have adopted the numerical approach to obtain the solution of the wave equation which leads to the expression for the Brans-Dicke scalar field $\Phi$. To do this, we chose the initial condition $\Phi(0)=\Phi_0=\text{constant}$ and $\Phi^{\prime} (0)=0$. Before embarking on a discussion of the physical viability of our model we like to comment on the magnitude of the BD coupling constant, $\omega_{BD}$.  A review of the literature reveals that lower bound on the Brans-Dicke parameter is $|\omega_{BD}| > 40,000$. This bound arises from the Doppler tracking data of the
Cassini spacecraft. However, the shortcoming of such experiments is that they are restricted to ‘‘weak-field’’
limits and cannot reveal spatial or temporal deviations of
the gravitational constant on significantly larger scales. First year data from WMAP restricted the BD constant to $\omega_{BD} > 1000$ with further refinements coming through an array of observational data pinning $\omega_{BD} > 120$. A more recent attempt in refining the bounds on $\omega_{BD}$ uses data from CMB (WMAP 5 yr, ACBAR 2007, CBI polarization, the BOOMERANG 2003 ) and the LSS data (galaxy clustering power spectrum from SDSS DR4 LRG data). This puts a restriction of $\omega_{BD} < - 120.0$ or $\omega_{BD} > 97.8$ which differs in at least two orders of magnitude from  $|\omega_{BD}| > 40000$ \cite{feng1,feng2}. The range of $\omega_{BD}$ (5; 20) chosen in this work is in agreement with the solar system constraints of the Brans-Dicke (BD) gravity presented by Perivolaropoulos \cite{peri}, where $\omega_{BD} \gtrsim -3/2$ for all $m_\phi \gtrsim 2 \times 10^{-6} \equiv 2 \times 10^{-16} ~eV$. The mass of the scalar particle $m_\phi$ is taken to be $0.002 \equiv 2 \times 10^{-13}~eV$ and is within the predicted range for neutron stars according to Popchev et al. \cite{pop} i.e. $10^{-16} ~eV \lesssim m_\phi \lesssim 10^{-9} ~eV$ or in dimensionless unit  $10^{-6} \lesssim m_\phi \lesssim 10$. Therefore, the range of Brans-Dicke parameter $\omega_{BD}$ used in this work is physically motivated for the chosen scalar particle mass $m_\phi =0.002$.

\subsection{Central values of the physical parameters}

In the interior of the compact star, the central values of all the physical parameters must be finite and non-singular. Firstly, we find the central values for both metric functions at $r=0$ as: $e^{\eta(0)}=1$ and $\xi(r)=2\ln[A+\sqrt{a}B/b]$. {Since $r=0$ is a zero for the function $\Phi^{\prime}(r)$, then by the factor theorem we can write: $\Phi^{\prime}(r)=r\,\psi(r)$, where $\psi(r)$ is any function of $r$. We then obtain values for density and pressures at the centre $r=0$
\begin{eqnarray}
\rho(0) &=& \frac{6\, b \,C\, \Phi_0 + 6\, a\, (A + C)\, \Phi_0 -3\, (A + C)\, \psi_0}{32\,\pi\,(A + C)}+\mathcal{B}\\
p_r(0) &=&\frac{2\,b\, C \, \Phi_0 + 2\, a\, (A + C)\, \Phi_0 - (A + C) \psi_0}{32\,\pi\,(A + C)}-\mathcal{B}\\
p_t(0) &=& \frac{14\, b \,C\, \Phi_0 - 
 10\, a\, (A + C) \Phi_0 - (A + C)\, (2\, m^2_0 \,\Phi^2_0 - 17 \psi_0)}{32\,\pi\,(A + C)}+\mathcal{B}.~~~~
\end{eqnarray}
Here we observe that $p_r(0) \neq p_t(0)$ which leads to $\Delta(0) \neq 0$. This situation may create certain problems in stellar modeling as the TOV will be unable to balance at the center. In order to nullify the anisotropy at $r=0$, we must have $p_r(0)=p_t(0)$, which leads to 
\begin{eqnarray}
b &=& \frac{(A + C) (6\,a \,\Phi_0 + m^2_{\phi}\,\Phi^2_0 - 32\,\mathcal{B}\, \pi-9\, \psi_0)}{6\,C\,\Phi_0}. \label{e5.4}
\end{eqnarray}
Now the central value of $E^2$ is given by
\begin{eqnarray}
E^2(0) &=& \frac{-6\, b\, C\,\Phi_0 + 
 6\,a\,(A + C)\,\Phi_0 + (A + C)\,(m^2_{\phi}\,\Phi^2_0 - 32\,\mathcal{B}\,\pi - 9\,\psi_0)}{(A + C)}. \label{e5.5}
\end{eqnarray}
As we can see from above Eq.(\ref{e5.5}) that $E^2(0)$ is nonvanishing. However, incorporating Eq. (\ref{e5.4}) into Eq. (\ref{e5.5}), we obtain $E^2(0)=0$, as required. We will now discuss the regularity of these physical parameters throughout the stellar interior.}

\subsection{Regularity}
\begin{figure*}[thbp]
\centering
\includegraphics[width=7.5cm]{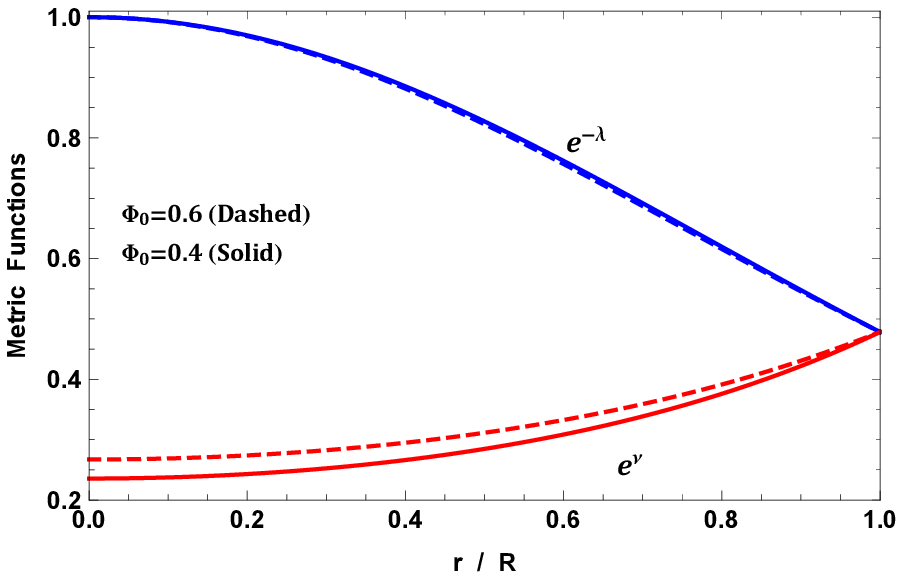}
\includegraphics[width=7.5cm]{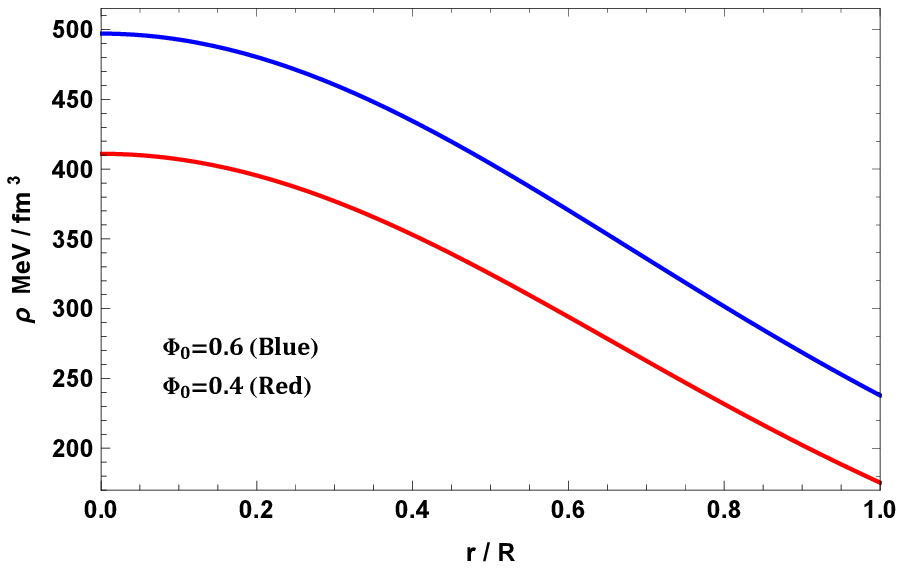}
\caption{The metric potentials and matter density for PSR J1903+327  plotted against $r/R$ by taking
$\Phi_0=0.4,\omega = 5; m_\phi = 0.002; B = 0.0457; \beta = -0.0088; b = 0.004/km^2; \alpha = 0.397, \mathcal{B}=56.998~MeV/fm^3$ and $\Phi_0=0.6,\omega = 5; m_\phi = 0.002; B = 0.0383; \beta = -0.0109; b = 0.0035/km^2; \alpha = 0.596, \mathcal{B}=79.237~MeV/fm^3$ }\label{f1}
\end{figure*}

\noindent (i) Metric functions at the centre, $r=0$: we observe from Fig \ref{f1} (left panel) that the metric functions at the centre $r=0$ assume finite values and are smooth and continuous throughout the interior of the stellar configuration. We conclude that the 
metric functions are free from singularity and positive at the centre.
\\

\noindent (ii) The density as a function as the radial coordinate is displayed in Fig. \ref{f1} (right panel). We observe that the density assumes a finite value at the centre and decreases monotonically towards the stellar surface. Furthermore, we observe that a higher value for $\Phi_0$ leads to a higher density at each point of the stellar interior.\\

\begin{figure*}[thbp]
\centering
\includegraphics[width=7.5cm]{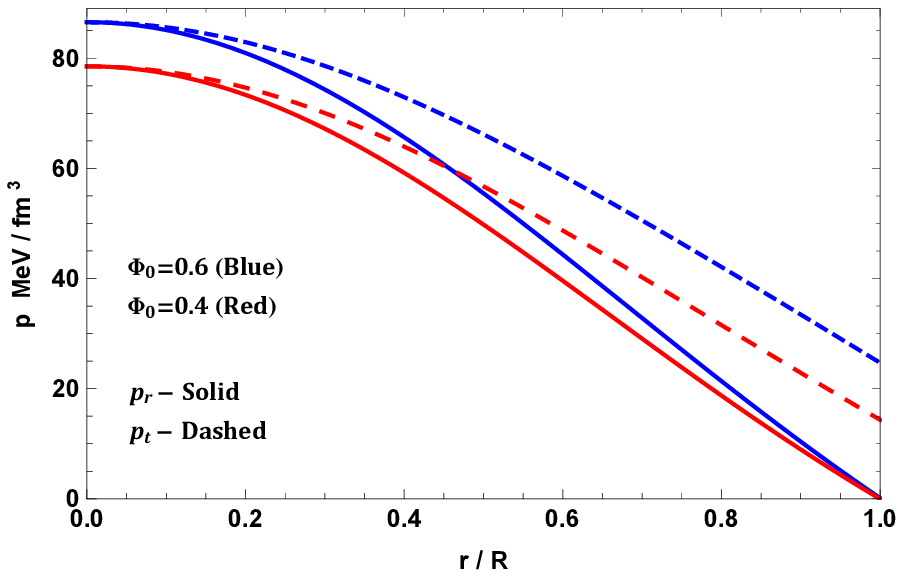}
\includegraphics[width=7.5cm]{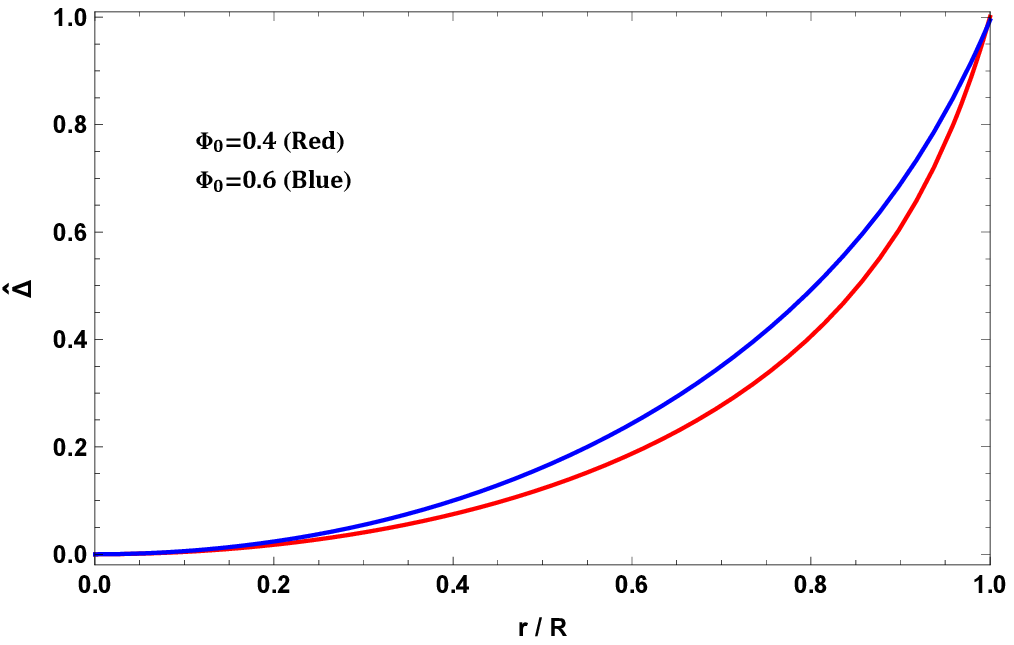}
\caption{The pressure and $\hat{\Delta}=(p_t-p_r)/p_t$ are plotted against $r/R$ by taking the same values as in Fig. \ref{f1} for PSR J1903+327. }\label{f2}
\end{figure*}

\noindent (iii) Pressure at the centre $r=0$: Fig. \ref{f2}  (left panel) shows us that both the radial and tangential pressure are regular at the centre of the star and decrease smoothly towards the boundary. The vanishing of the radial pressure at some finite value, $r = r_\Sigma$ defines the boundary of the fluid sphere. From the equation of state and the density profile we note that an increase in the magnitude of $\Phi_0$ leads to an increase in both the radial and transverse pressures. It is widely acknowledged that while the radial pressure vanishes at the boundary the tangential pressure need not. A phenomenological explanation is provided in the work by Boonserm et al.\cite{boon}. They point out that if the tangential pressure does not vanish at the boundary of the star then matching to the Schwarzschild exterior would imply that the electric field is discontinuous at the stellar surface. This would point to the existence of a nonzero surface charge density. To avoid this peculiarity the interior spacetime can be matched to a Reissner-Nordstr{\"o}m exterior implying that the stellar object is electrically charged. The nonzero transverse pressure at the boundary is then  balanced by the transverse stress of the electric field in the exterior. \\

\noindent (iv) The anisotropy parameter is presented in Fig. \ref{f2} (right panel) and it is clear that $\Delta > 0$ at each interior point of the stellar configuration. The anisotropy parameter vanishes at the center of the star and increases to a maximum for some finite radius, $r < r_{\Sigma}$ where ${\Sigma}$ denotes the boundary of the star. A positive value for $\Delta$ ($p_t > p_r$) signifies a repulsive force due to anisotropy. It is clear that the increase in the anisotropy parameter especially towards the surface layers lead to greater stability in these regions. It is interesting to observe that a larger value for $\Phi_0$ leads to increased values for the anisotropy parameter thus stabilising the fluid even further. 

\subsection{Equation of state}

The role of the equation of state (EoS) has been demonstrated in many models of compact objects within the framework of classical general relativity and modified theories of gravitation. A barotropic EoS of the form $p = p(\rho)$ points strongly to the type of matter making up the star. Recently, the colour-flavoured locked-in EoS was utilised to model compact objects. This particular EoS is a generalisation of the MIT bag model and attempts to connect the microphysics to macrophysics of the fluid configuration. Fig. \ref{f3} (left panel) shows the variation of the ratio $p/\rho$ with $r/R$. We note that the pressure is less than the density at each interior point of the configuration. This ratio is also positive everywhere inside the star.

\subsection{Electric field}

\begin{figure*}[thbp]
\centering
\includegraphics[width=7.5cm]{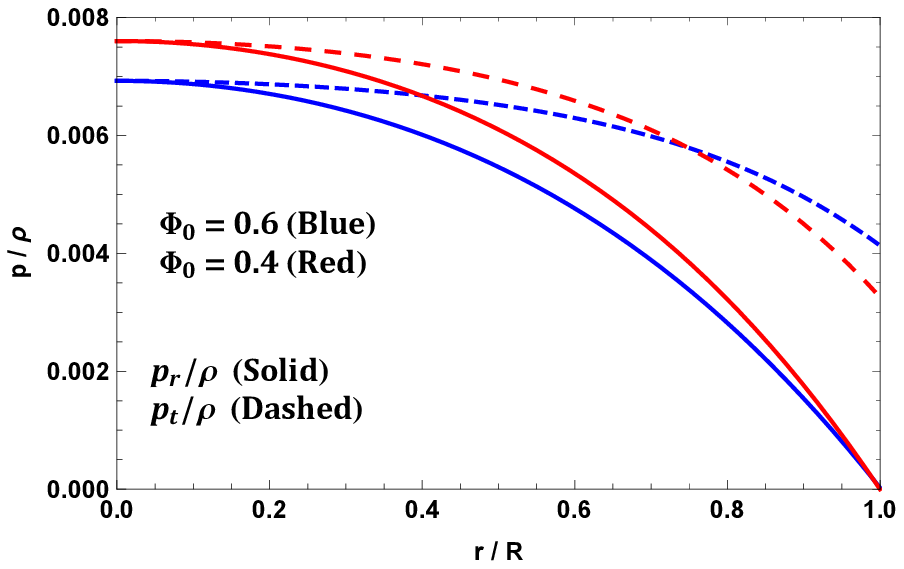}
\includegraphics[width=7.5cm]{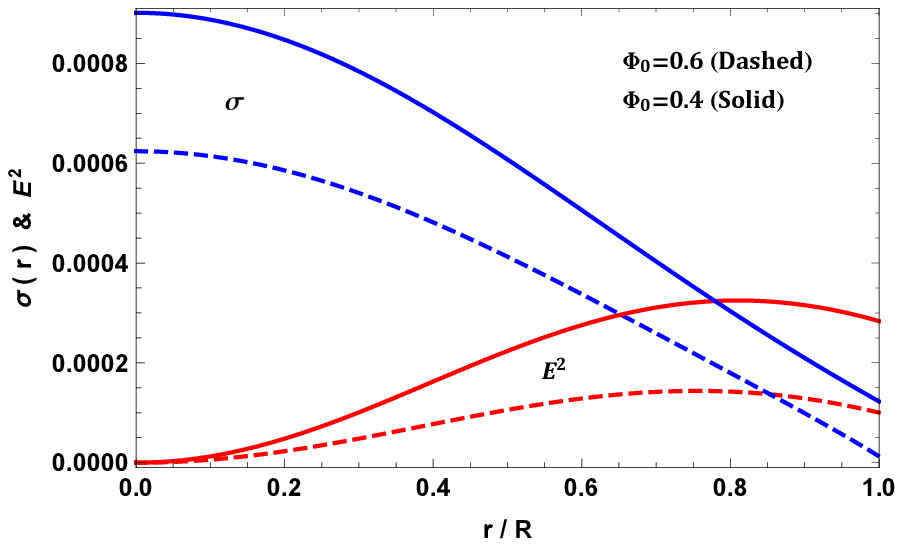}
\caption{The $p/\rho$ and electric field intensity are plotted against $r/R$ by taking the same values as in Fig. \ref{f1} for PSR J1903+327. }\label{f3}
\end{figure*}

The trend in the electric field and the charge density are shown in Fig. \ref{f3} (right panel). We observe that the electric field vanishes at the centre of the star and increases monotonically towards the surface. It is well-known that intense electric fields can lead to instabilities within the stellar core. The presence of charge as high as $10^{20}$ coulombs can generate quasi-static equilibrium states. These high charge densities are linked to very intense electric fields which in turn induce pair production within the star thus leading to an unstable core. 

\subsection{Energy conditions}
In order for physical admissibility of our models the solution should satisfy the following energy conditions, viz., (i) null energy condition (NEC), (ii) weak energy
condition (WEC) and (iii) strong energy condition (SEC).
In order to satisfy the above energy conditions, the following inequalities must be hold simultaneously at each interior point of the  the charged fluid sphere:

\begin{eqnarray}
\text{NEC}: \rho+\frac{E^2}{8\pi}\geq 0,~~~\text{WEC}: \rho+p_i +\frac{E^2}{8\pi} \geq  0,~~~
\text{SEC}: \rho+\sum_i p_i+\frac{E^2}{4\pi} \geq  0.\label{eq24}
\end{eqnarray}
It is clear from Fig. \ref{f4}. (left panel) that all three energy conditions are satisfied at each interior point of the configuration.
\subsection{TOV equation}
\begin{figure*}[thbp]
\centering
\includegraphics[width=7.5cm]{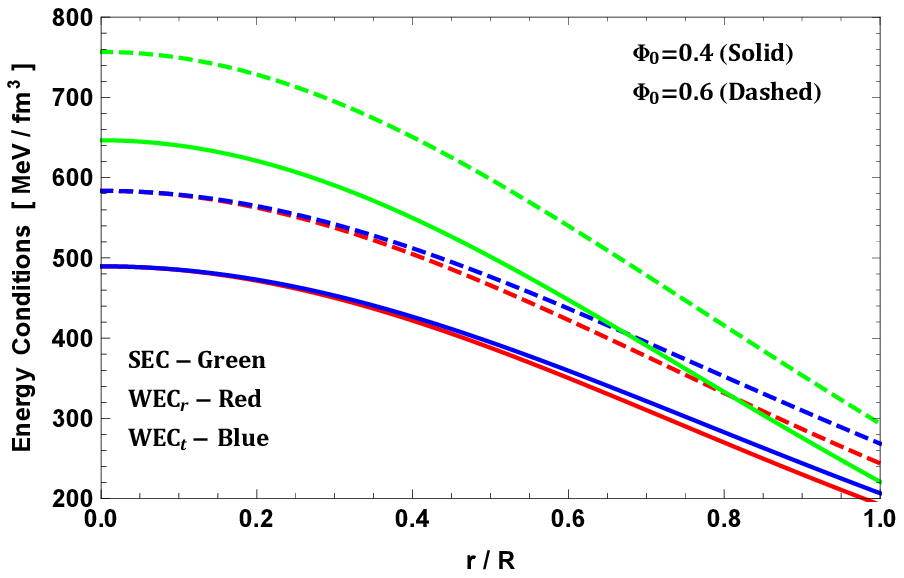}
\includegraphics[width=7.5cm]{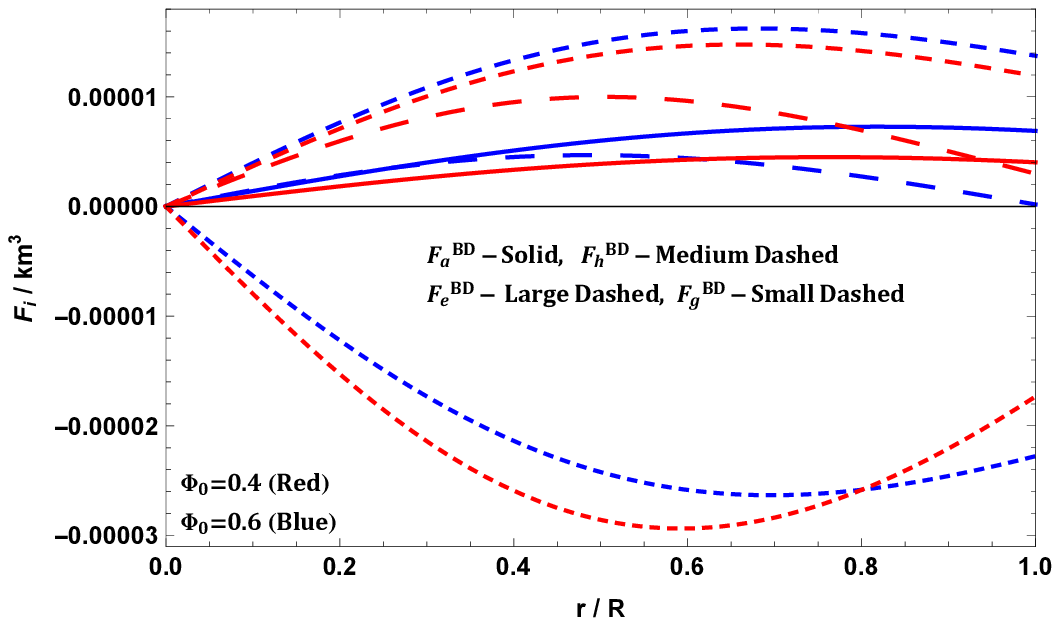}
\caption{The energy conditions and TOV-equation are plotted against $r/R$ and $r$ respectively by taking the same values as in Fig. \ref{f1} for PSR J1903+327. }\label{f4}
\end{figure*}

It is well known that in the absence of any dissipative effects such as heat flow the equilibrium of a 
gravitationally bounded charged fluid configuration is characterised by the resultant of the gravitational force, $F_g$, the
hydrostatic $F_h$ force, the force due to anisotropy, $F_a$ and the elctrostatic interaction, $F_e$ vanishing at each interior
point of the star. The modified TOV equation in Brans-Dicke gravity is given by 
\begin{eqnarray}
-p_r^{\prime}-\frac{\xi^{\prime}}{2}(p_r+\rho)+\frac{2}{r}(p_t-p_r)+\big(T^{1(\Phi)}_1\big)^\prime-\frac{\xi^{\prime}}{2}\big(T^{1\Phi}_1-T^{0\Phi}_0\big)  +\frac{2}{r}\big(T^{1\Phi}_1-T^{2\Phi}_2\big)+\frac{qq^{\prime}}{4\,\pi r^4}=0,~~~~ \label{e16}
\end{eqnarray}

where,
\begin{eqnarray} \\
&&F_h^{BD}=-p_r^\prime+\big(T^{1\Phi}_1\big)^\prime;~~F_g^{BD} =-\frac{\xi'}{2}\,\big(p_r+\rho+T^{1\Phi}_1-T^{0\Phi}_0\big);\nonumber\\&&F_a^{BD}=\frac{2}{r}\,\big(p_t-p_r+T^{1\Phi}_1-T^{2\Phi}_2\big);~~F_e^{BD} =\frac{qq^{\prime}}{4\,\pi r^4}\nonumber
\end{eqnarray}
We note the contribution of the scalar field to the hydrostatic, gravitational and anisotropic forces respectively. In a recent study, Herrera \cite{herpi} pointed out an interesting observation regarding the nonappearance of the tangential pressure in the gravitational force term. In the case of $p_t > p_r$, anisotropic spheres are more compact than their isotropic counterparts. In Fig. \ref{f4} (right panel), we can see that the combined forces of electric, hydrostatic and anisotropic counter-balanced the gravity so that the configuration is under equilibrium. It can also be seen that when increasing the scalar field contribution (by increasing the initial condition of the Brans-Dicke scalar field $\Phi_0$) all the forces also increase proportionally. This contribution from the scalar field is a mechanism which enables the system to support a larger mass within each concentric shell i.e. the equation of state will be stiffened.

\subsection{Causality}
In order to prevent superluminal speeds within the stellar fluid we require that the speed of sound be less than the speed of light everywhere inside the star. The speed of sound for the charged fluid sphere should be monotonically decreasing from centre to the boundary of the star ($v=\sqrt{dp/d\rho} < 1$). It is clear from Fig. \ref{f5} (left panel) that the speed of sound is less than $1$ throughout the interior of the matter distribution. This implies that our fluid model fulfills causality requirements throughout its interior.

\subsection{Stability factor}
\begin{figure*}[thbp]
\centering
\includegraphics[width=7.5cm]{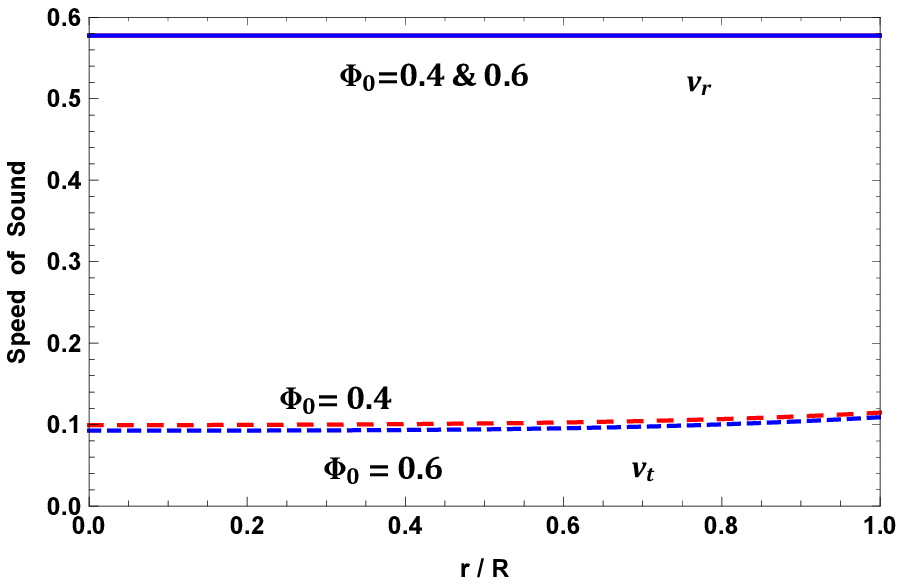}
\includegraphics[width=7.5cm]{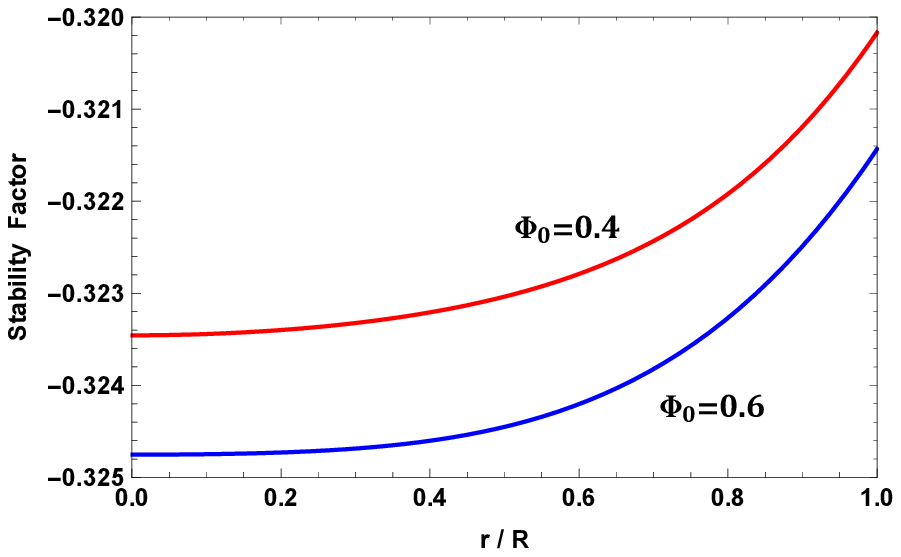}
\caption{The speed of sound and stability factor are plotted against $r/R$ by taking the same values as in Fig. \ref{f1} for PSR J1903+327. }\label{f5}
\end{figure*}

We observe from Fig. \ref{f5} (right panel) that our model satisfies the Herrera cracking condition $-1 <v_t^2-v_r^2<0$. Abreu et al. \cite{Ab}  showed that stable and unstable patches can arise within the stellar fluid and their existence depends on the relative sound speeds in the radial and tangential directions. In particular, potential unstable regions occur when the tangential component of the speed of sound exceeds the radial component.

\subsection{Stability through adiabatic index}
 Within the Newtonian formalism of gravitation, it is also well known that there has no upper mass limit if the EoS has an adiabatic index $\Gamma > 4/3$ where

\begin{equation}
\Gamma=\frac{p+\rho}{p}\,\frac{dp}{d\rho}\, \label{eq31}
\end{equation}
the definition of which arises from an assumption within the Harrison-Wheeler formalism \cite{Harrison}. A perturbative study of dissipative collapse by Chan et al. \cite{channy} in which gravitational collapse proceeds from an initially static configuration  Eq. (\ref{eq31}) follows from the EoS of the unperturbed, static matter distribution.  Eq. (\ref{eq31}) is modified in the presence anisotropic fluids (radial and transverse stresses are unequal) and we can write  
\begin{equation} \label{eqgamma}
\Gamma > \frac{4}{3} - \left[\frac{4}{3}\frac{p_r - p_t}{rp_r'}\right]_{max}\end{equation}
It is well-known that a bounded {\em charged} configuration can be treated as an anisotropic system. In the special case of isotropic pressure ($p_r = p_t$) the classical Newtonian result holds from Eq. (\ref{eqgamma}). Observations of Eq. (\ref{eqgamma}) indicate that instability is increased when $p_r < p_t$ and decreases when $p_r > p_t$. Fig. \ref{f6} (left panel) confirms that our model is stable against radial perturbations at each interior point within the stellar fluid. 
\subsection{Stability through Mass-central density ($M-\rho_c)$ curve}
\begin{figure*}[thbp]
\centering
\includegraphics[width=7.5cm]{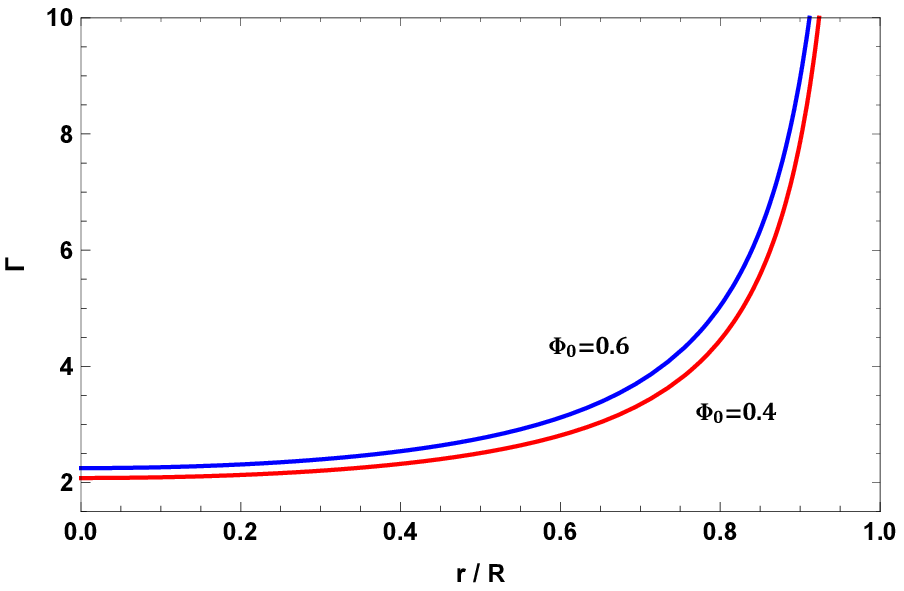}
\includegraphics[width=7.5cm]{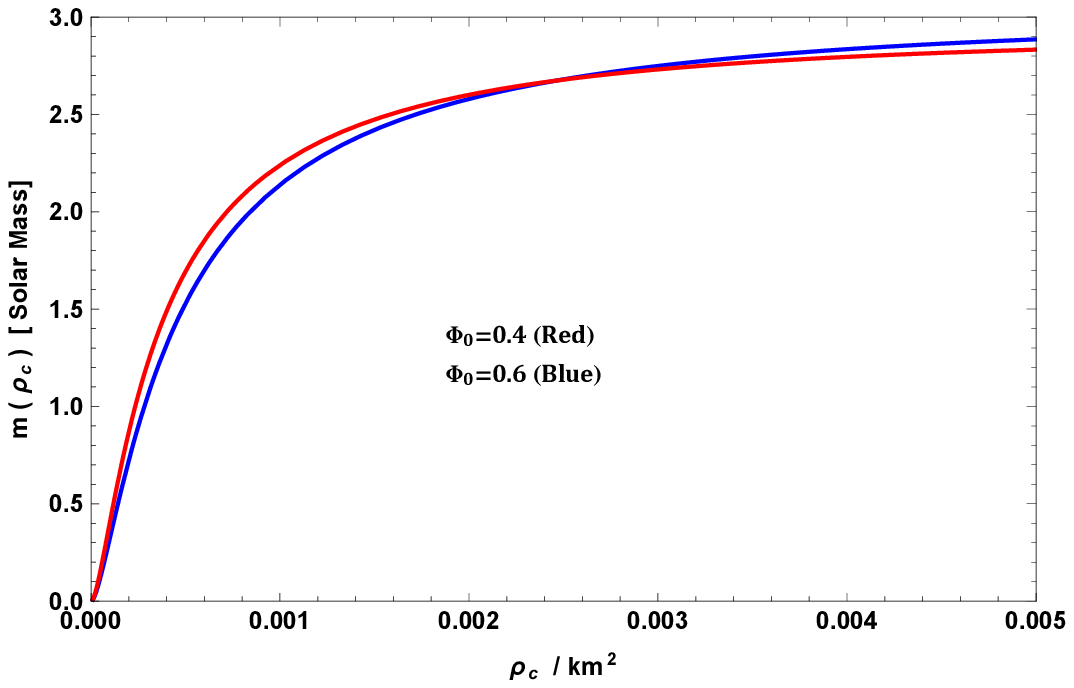}
\caption{The adiabatic index and mass-central density curves are plotted against $r/R$ and $\rho_c$ by taking the same values as in Fig. \ref{f1} for PSR J1903+327. }\label{f6}
\end{figure*}

Now, we focus on the $M-\rho_c$ function dubbed as static stability criterion which is a noteworthy thermodynamically quantity in order to give more insight into the stability of the compact celestial structure. This static stability criterion has been developed and made more accessible by Harrison and co-workers \cite{H66} and Zeldovich $\&$ Novikov \cite{Z71} after suggestions by Chandrashekhar \cite{C64} in order to portray the stability of gaseous celestial configuration according to radial pulsations. In this respect, the formula associated between the gravitational mass $M$ and the central density $\rho_c$ is given as follows,
\begin{equation}
\frac{\partial M(\rho_c)}{\partial \rho_c} > 0,
\end{equation}
which must be satisfied in order to describe the solutions of static and stable celestial configurations or otherwise unstable if
\begin{equation}
\frac{\partial M(\rho_c)}{\partial \rho_c} < 0,
\end{equation}
under radial perturbation. We present the variation of the gravitational mass $M$ with respect to central density $\rho_c$ in Fig. \ref{f6} (right panel). It shows that in the present study the gravitational mass $M$ is an increasing function with regard to central density for both initial condition values of the BD scalar field $\Phi_0$ i.e., $\Phi_0=0.4$ and $\Phi_0=0.6$, by tuning $\omega_{BD}$ to $5$. This confirms the static stability criterion of the stellar system against radial perturbations. We can see that the solution takes its stability with the increase of the initial condition parametric values of $\Phi_0$ and we also find that the celestial bodies become more massive according to increasing central density.

\subsection{Effective mass and compactness parameter for the charged compact star}
As a starting point we recall that the maximal absolute limit of mass-to-radius $(M/R)$ ratio for a static
spherically symmetric isotropic fluid model is given by $2M/R\leq 8/9 $ \cite{Bu}. In the case of charged fluid spheres ~\cite{Boehmer2006} showed that there exists a lower bound for the mass-radius ratio
\begin{equation}
\frac{Q^{2}\, (18 R^2+ Q^2) }{2R^{2}\, (12R^2+Q^2)}  \leq
\frac{M}{R}, \label{eq62}
\end{equation}
for the constraint $Q < M$.

However this upper bound of the mass-radius ratio for charged compact star was
generalized by~\cite{And} who proved
that
\begin{equation}
\frac{M}{R} \leq \left[\frac{2R^2+3Q^2}{9R^2} +\frac{2}{9R}\,\sqrt{R^2+3Q^2}\right]. \label{eq63}
\end{equation}

The  Eqs. \ref{eq62} and \ref{eq63} imply that
\begin{equation}
\frac{Q^{2}\, (18 R^2+ Q^2) }{2R^{2}\, (12R^2+Q^2)}  \leq
\frac{M}{R} \leq \left[\frac{2R^2+3Q^2}{9R^2} +\frac{2}{9R}\,\sqrt{R^2+3Q^2}\right]
\end{equation}

The effective mass of the charged fluid sphere can be determined as:
\begin{equation}
m_{eff}=4\pi{\int^R_0{\left(\rho+\frac{E^2}{8\,\pi}\right)\,r^2\,dr}}=\frac{R}{2} \Big[1-e^{-\eta(R)}\Big]\, \label{eq32}
\end{equation}
where $e^{-\eta}$ is given by the equation (\ref{e4.1}) and compactness $u(r)$ is defined as:

\begin{equation}
u(R)=\frac{m_{eff}(R)}{R}=\frac{1}{2} \Big[1-e^{-\eta(R)} \Big]\, \label{eq33}
\end{equation}

\subsection{Redshift}

The maximum possible surface redshift for a bounded configuration with isotropic pressure is $Z_s = 4.77$. In the work of Bowers and Liang they showed that this upper bound can be exceeded when the radial and transverse pressures are different\cite{bowers}. In particular, when $\Delta > 0$ ($p_t > p_r$) the surface redshift is greater than its isotropic counterpart.  Studies show that for strange quark stars the surface redshift is higher in low mass stars with the difference being as high as 30$\%$ for a 0.5 solar mass star and 15$\%$ for a 1.4 solar mass star. It appears that higher redshift predictions in low mass stars appear to be an anomaly. In a recent study Chandra et al. \cite{chandra1} have used gravitational redshift measurements to determine the mass-radius ratio of white dwarfs. Using data of over three thousand catalogued white dwarfs they were able to determine the mass-radius relation over a wide range of stellar masses. Their improved technique entailed the cancelling of random Doppler shifts by averaging out the apparent radial velocities of white dwarfs with similar radii enabling them to measure the associated gravitational redshift.

The gravitational surface red-shift ($Z_s$) is given as:

\begin{equation}
Z_s= (1-2\,u)^{-1/2} -1, \label{zs}
\end{equation}

\begin{figure*}[thbp]
\centering
\includegraphics[width=7.5cm]{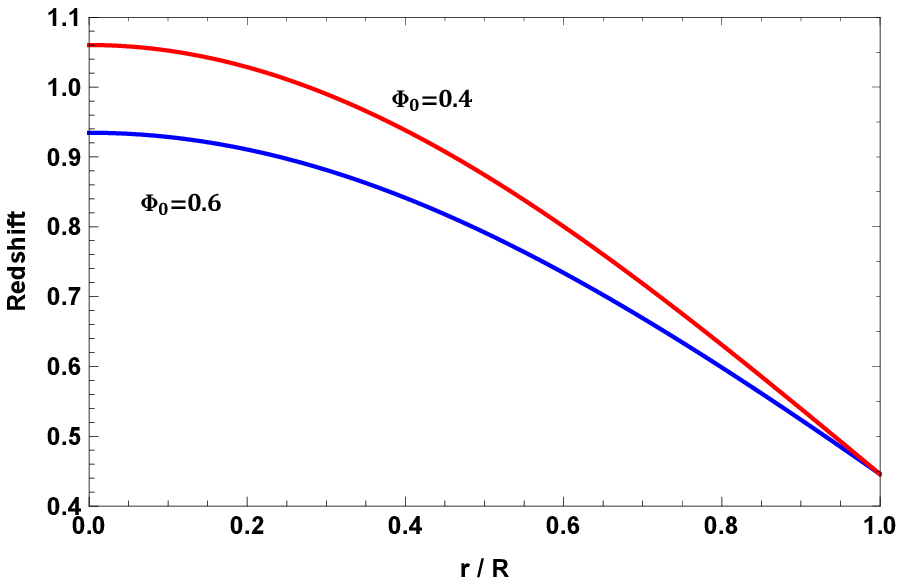}
\includegraphics[width=7.5cm]{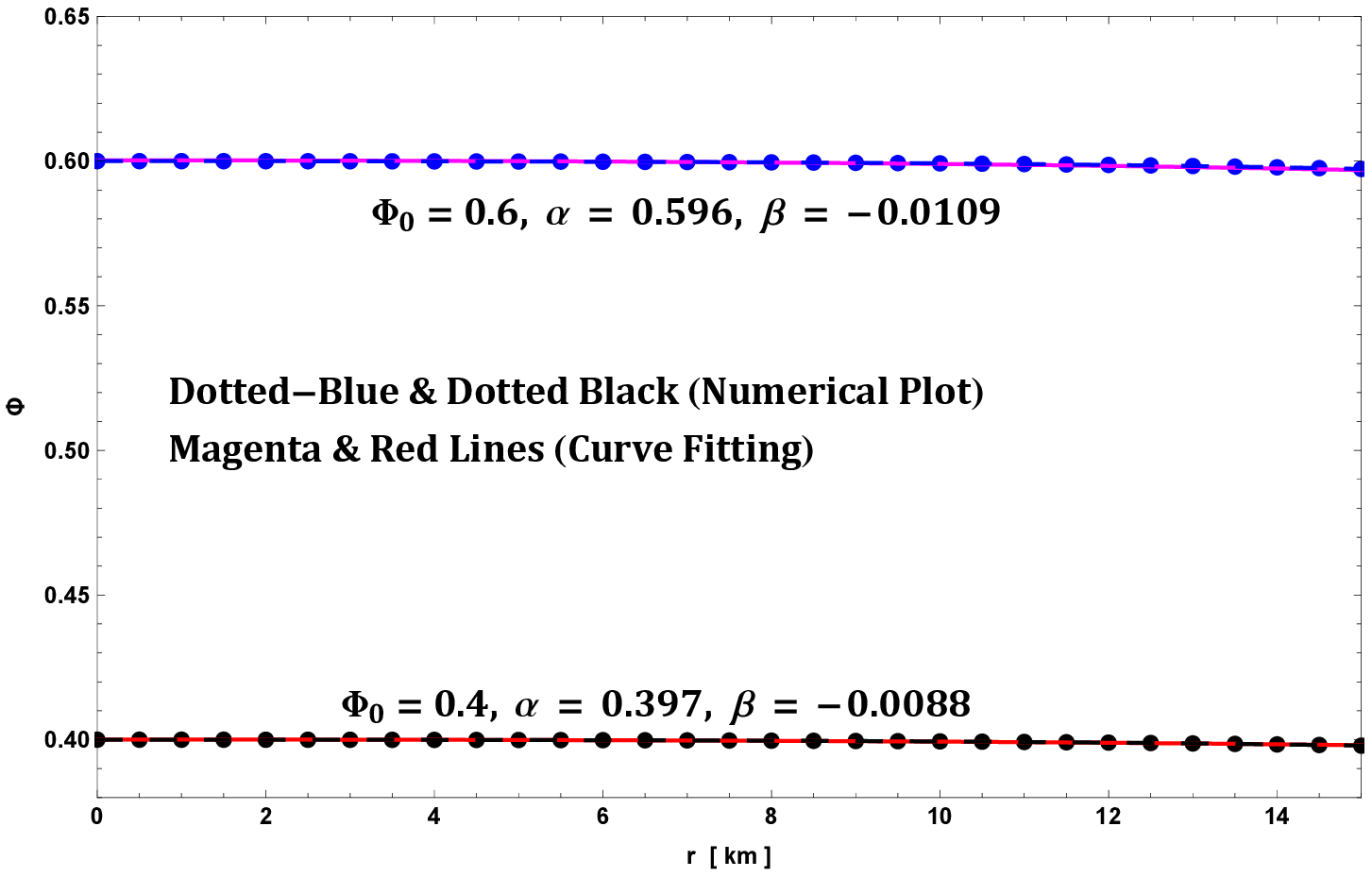}
\caption{The gravitational redshift profile and scalar field $\Phi(r)$ are plotted against $r/R$ by taking the same values as in Fig. \ref{f1} for PSR J1903+327.}\label{f7}
\end{figure*}

From Eq.(\ref{zs}), we note that the surface redshift depends upon the compactness $u$, which implies that the surface redshift for any star cannot be arbitrarily large because compactness $u$ satisfies the Buchdhal maximal allowable mass-radius ratio. However, the value for the surface redshift for the different compact objects have been calculated when $\Phi_0=0.4$ and $\Phi_0=0.6$ respectively with $\omega_{BD}=5$ as follows:(i) 0.4856 and  0.5637 for PSR J1903+327, (ii) 0.4446 and 0.4955 for Cen X-3, (iii) 0.4086 and 0.4489 for EXO 1785-248, (iv) 0.3690 and 0.3916 for LMC X-4. The graphical behavior for gravitational redshift for PSR J1903+327 is shown by Fig.(\ref{f7}) (left panel). Furthermore, the behavior of the scalar field $\Phi(r)$ with respect to $r$ is shown in Fig.(\ref{f7}) (right panel). From this graph, it is clear that the functional form of the scalar field $\Phi(r)$ is well-fitted with numerical data which also approves the validity of our new solutions.

\begin{table*}[thbp]
\centering
\caption{\label{table1}{Comparative study of lower bound, Mass-radius ratio, upper bound, compactness $(u=M_{\text{eff}}/R)$ and surface red-shift of the star for fix values of $\alpha=0.397,\beta=-0.0088,\Phi_0=0.4$ and $\omega_{BD}=5$.}}
\begin{tabular}{cccccccc}\hline
Objects & $Q(R)$  & Lower bound  & Mass-radius & Upper bound & $z_s$ \\
 & $\times 10^{19}$C & $\frac{Q^2\,(18R^2+Q^2)}{2R^2\,(12R^2+Q^2)}$ & ratio ($M/R$)&$\frac{2R^2+3Q^2+2R\sqrt{R^2+3Q^2}}{9R^2}$ &  \\
\hline
PSR J1903+327 & 1.743  & 0.0189 & 0.2734 & 0.461 & 0.4856\\
Cen X-3 & 1.648 & 0.0169 & 0.2604 & 0.459 & 0.4446\\
EXO 1785-248 & 1.532 & 0.0146 & 0.2480 & 0.457 & 0.4086\\
LMC X-4 & 1.349 & 0.0113 & 0.2330 & 0.455 & 0.3690\\
\hline
\end{tabular}
\end{table*} 
\begin{table*}[thbp]
\centering
\caption{\label{table2}{Comparative study of lower bound, Mass-radius ratio, upper bound, compactness $(u=M_{\text{eff}}/R)$ and surface red-shift of the star for fix values of $\alpha=0.596,\beta=-0.010888,\Phi_0=0.6$ and $\omega_{BD}=5$.}}
\begin{tabular}{cccccccc}\hline
Objects & $Q(R)$  & Lower bound  & Mass-radius & Upper bound & $z_s$ \\
 & $\times 10^{19}$C & $\frac{Q^2\,(18R^2+Q^2)}{2R^2\,(12R^2+Q^2)}$ & ratio ($M/R$)&$\frac{2R^2+3Q^2+2R\sqrt{R^2+3Q^2}}{9R^2}$ & \\
\hline
PSR J1903+327 & 2.889 & 0.0597 & 0.2955 & 0.4896 & 0.5637\\
Cen X-3 & 2.535 & 0.0400 & 0.2760 & 0.4794 & 0.4955\\
EXO 1785-248 & 2.214 & 0.0333 & 0.262 & 0.4737 & 0.4489\\
LMC X-4 & 1.808 & 0.0204 & 0.2420 & 0.4624 & 0.3916\\
\hline
\end{tabular}
\end{table*} 
\begin{table*}[thbp]
\centering
\caption{\label{table3}{ Comparative study of lower bound, Mass-radius ratio, upper bound, compactness $(u=M_{\text{eff}}/R)$ and surface red-shift of the star for fix values of $\alpha=0.596,\beta=-0.010888,\Phi_0=0.6$ and $\omega_{BD}=50$.}}
\begin{tabular}{cccccccc}\hline
Objects & $Q(R)$  & Lower bound  & Mass-radius & Upper bound & $z_s$ \\
 & $\times 10^{19}$C & $\frac{Q^2\,(18R^2+Q^2)}{2R^2\,(12R^2+Q^2)}$ & ratio ($M/R$)&$\frac{2R^2+3Q^2+2R\sqrt{R^2+3Q^2}}{9R^2}$ &\\
\hline
PSR J1903+327 & 2.8980 & 0.0523 & 0.2957 & 0.4899 & 0.5645\\
Cen X-3 & 2.5420 & 0.0402 & 0.2766 & 0.4796 & 0.4960\\
EXO 1785-248 & 2.2195 & 0.0307 & 0.2601 & 0.4714 & 0.4436\\
LMC X-4 & 1.8715 & 0.0218 & 0.2371 & 0.4637 & 0.3790\\
\hline
\end{tabular}
\end{table*} 

\subsection{The effect of BD scalar field $\Phi$, and BD-parameter $\omega_{BD}$ on the $M-R$ and $M-I$ curves}
\begin{figure*}[thbp]
\centering
\includegraphics[width=7.5cm]{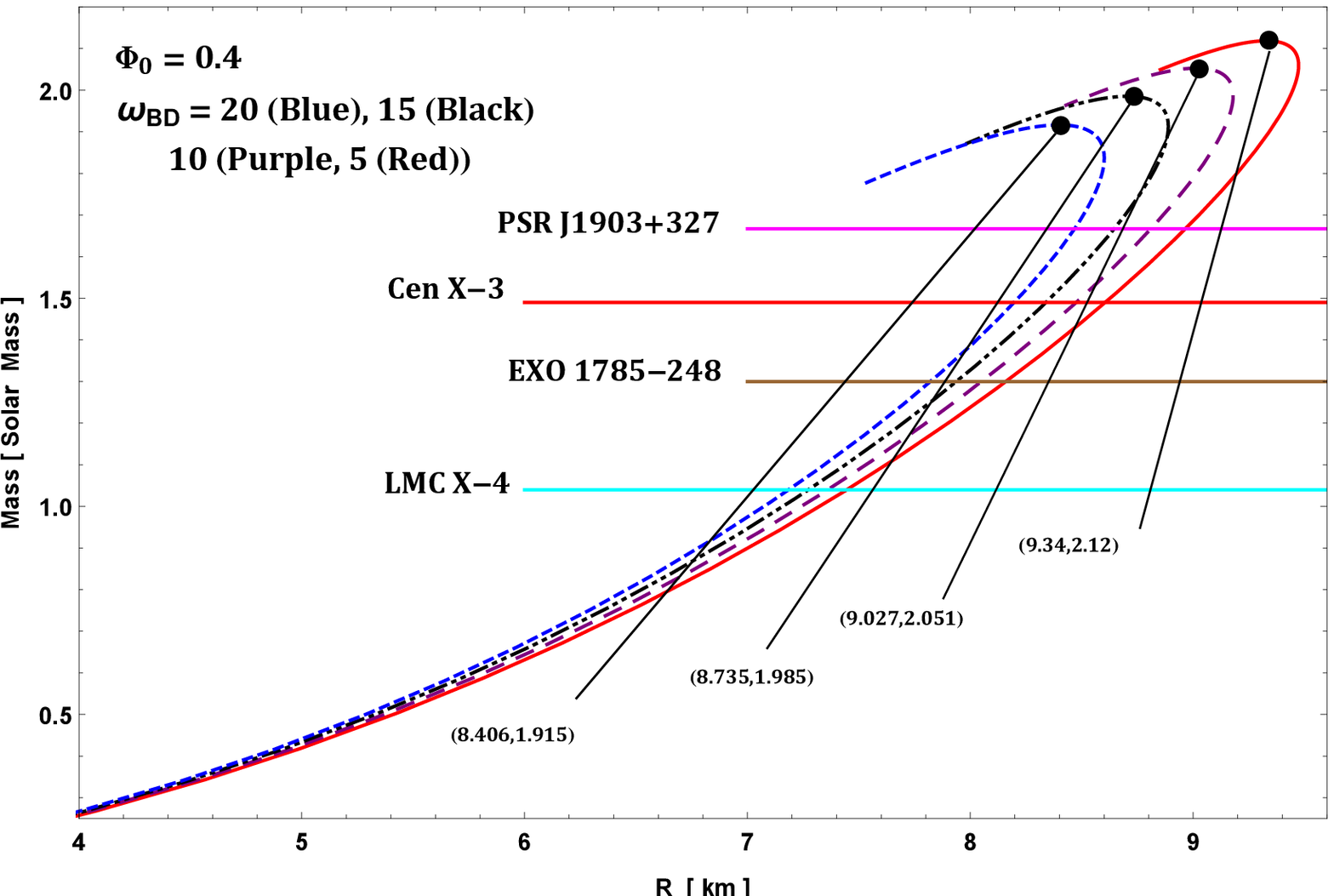}
\includegraphics[width=7.5cm]{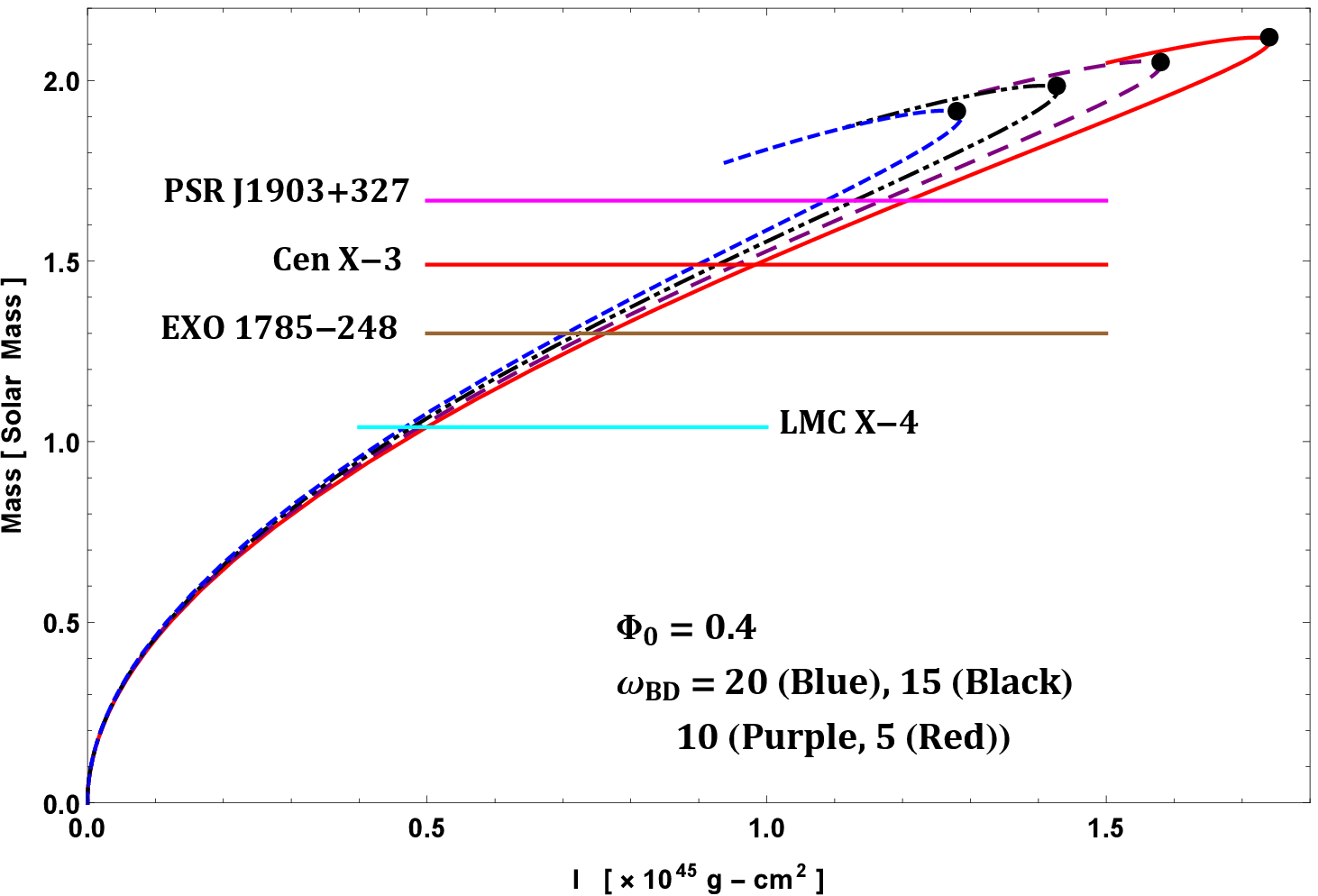}
\caption{The $M-R$ and $M-I$ curves are plotted by taking the same values as in Fig. \ref{f1} for different values of $\omega_{BD}$. }\label{f8}
\end{figure*}

\begin{figure*}[thbp]
\centering
\includegraphics[width=7.5cm]{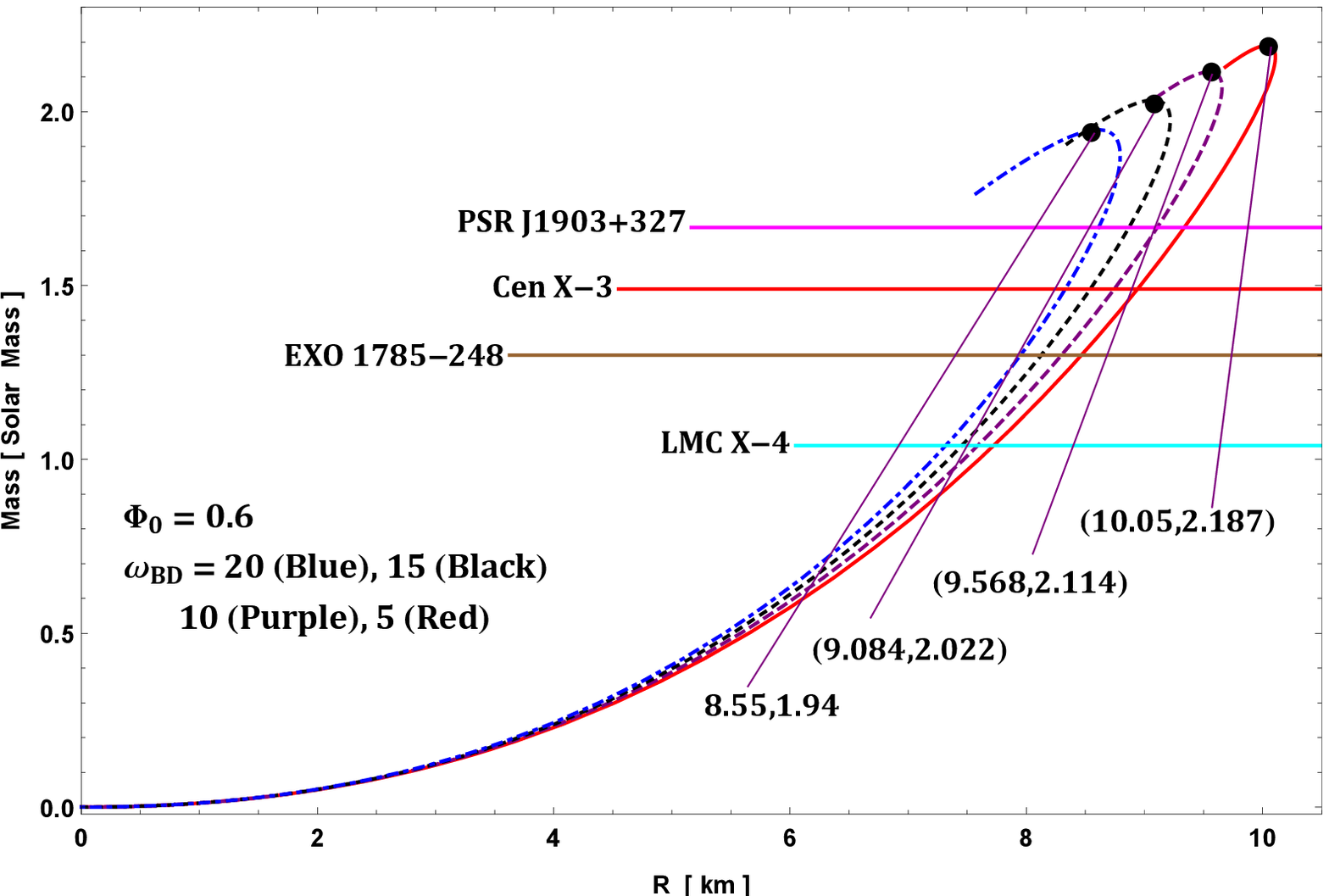}
\includegraphics[width=7.5cm]{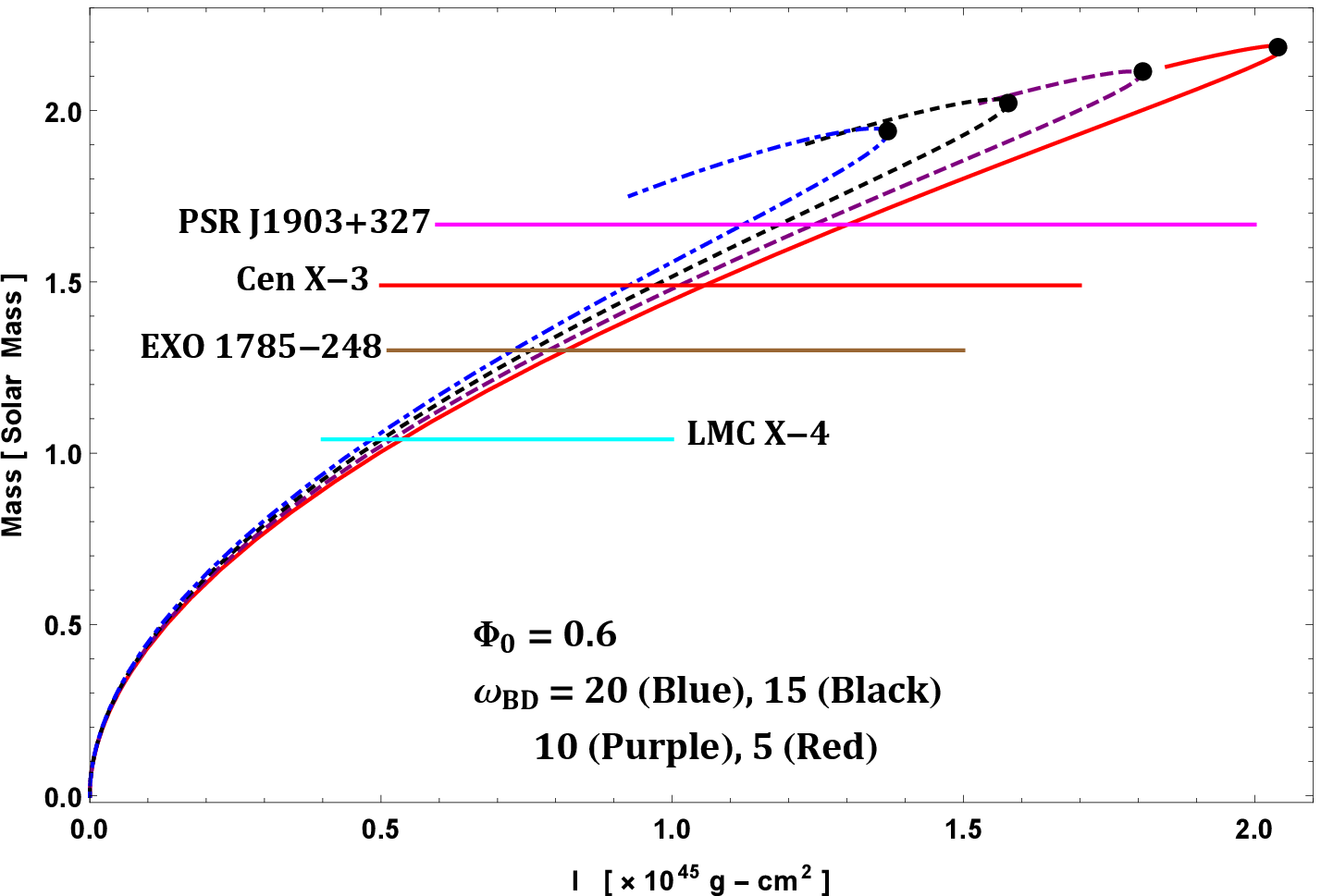}
\caption{The $M-R$ and $M-I$ curves are plotted against $r$ by taking the same values as in Fig. \ref{f1} for different values of $\omega_{BD}$. }\label{f9}
\end{figure*}

In this section, we examine the $M-R$ and $M-I$ diagrams resulted from our stellar model in the background of BD gravity with a massive field via the embedding approach. In this respect, we provide an instructive explanation of the influences included by the choices made on different parameters viz., the initial condition of the BD scalar field $\Phi_0$, BD-parameter $\omega_{BD}$ and the total external bag pressure $\mathcal{B}$ (or bag constant), in order to give a more achievable scenario and efficient astrophysical stellar system. On the other hand, for determining the stiffness of an EoS, we can analyze the moment of inertia $I$ associated with a static celestial solution which could give a precise instrument via adopting the Bejger $\&$ Haensel concept \cite{bh02}, given by,
\begin{equation}
I = {2 \over 5} \left(1+{(M/R)\cdot km \over M_\odot} \right)~MR^2.~~~~ \label{eq6.9}
\end{equation}
Our survey on $M-R$ and $M-I$ curves is highly significant for the stellar systems which clearly show the state (more or less) of compact celestial bodies via the maximum bound of the total mass as well as the efficacy and the sensitivity to the stiffness of an EoS. In this regard, from Figs. \ref{f8} and \ref{f9}, we show the variation of the total mass $M$ in [$M_{\odot}$] versus the total radial coordinate $R$ in [km] and the maximum moment of inertia $I$ in [$\times 10^{45}~g-cm^2$], for all chosen values of the parameters $\Phi_0$, $\omega_{BD}$ and $\mathcal{B}$. In the present BD gravity stellar model via the embedding approach, we observe  from $M-R$ curves featured in Fig. \ref{f8} (left panel) the contributions due to $\omega_{BD}$ where we have set $\Phi_0$ to $0.4$ and $\mathcal{B}$ to $56.998~MeV/fm^3$. We note that as the parameter $\omega_{BD}$ decreases from $20$ to $5$, the most extreme value of mass $M$ increases with an increase in the total radial coordinate $R$, which culminates in more massive compact celestial bodies. Moreover, from the $M-R$ curves illustrated in Fig. \ref{f9} (left panel) we further observe the effect of $\omega_{BD}$ by increasing $\Phi_0$ to $0.6$ with $\mathcal{B}=79.237 MeV/fm^3$. We can conclude that as $\omega_{BD}$ increases the corresponding radius $R$ decreases and the most extreme value of mass $M$ also decreases, which gives us also a celestial system less compact and less massive. This shows that the parameters $\Phi_0$, $\mathcal{B}$ and $\omega_{BD}$ will affect the maximum mass limit as well as compactness of the objects. On the other hand, the variation of the maximum moment of inertia $I$ with respect to the total mass $M$ due to the impact of $\Phi_0$, $\mathcal{B}$ and $\omega_{BD}$ has been featured in Figs. \ref{f8} (right panel) and \ref{f9} (right panel). From these plots, we can see that the maximum moment of inertia $I$ is always increasing with increasing the mass until up to the most extreme value of mass and decreasing rapidly with decreasing the mass. Consequently, we can infer that the stiffness of EoS is better in the case where $\Phi_0=0.6$ and $\omega_{BD} = 5$ with respect to all other cases, i.e., when $\omega_{BD} = 10,~15,~20$ and $\Phi_0= 0.4,~0.6$. It is worth mentioning here that the Bag constant $\mathcal{B}$ is changing with only $\Phi_0$ therefore $\mathcal{B}$ will also feature in the $M-I$ curves. 
Finally, we would like to mention here that we have discovered a good agreement with observational data for four compact celestial objects namely, PSR J1903+327, Cen X-3, EXO 1785-248, LMC X-4 in our resulting $M-R$ and $M-I$ curves. It is clear that all parameters introduced by the BD gravity stellar model with the massive field via the embedding approach have a large effect on the various physical parameters of the celestial configuration.

\begin{table*}[thbp]
\centering
\caption{{ Predicted radii and MI for some compact stars for different values of $\omega$ with $\alpha=0.397,\beta=-0.0088,\Phi_0=0.4$ correspond to Fig. \ref{f8}.}}\label{tab4}
 \scalebox{0.95}{\begin{tabular}{| *{10}{c|} }
\hline
{Objects} & {${M \over M_\odot}$}   & \multicolumn{4}{c|}{{Predicted $R$ km}} & \multicolumn{4}{c|}{{$I \times 10^{45} \,g-cm^2$}}\\
\cline{3-10}
&& \multicolumn{4}{c|}{$\omega_{BD}$} & \multicolumn{4}{c|}{$\alpha$}\\
\cline{3-10}
&  & 20 & 15 & 15 & 5 & 20 & 15 & 10 & 5  \\ \hline
PSR J1903+327   &   1.667  &   8.470  &   8.650  &   8.807  &   8.974  &   1.08  &   1.126  &   1.169 & 1.209 \\
\hline
Cen X-3 & 1.49 & 8.189 & 8.338 & 8.478 & 8.611 & 0.902 & 0.923 & 0.959 & 0.991\\
\hline
EXO 1785-248 & 1.3 & 7.825 & 7.849 & 8.073 & 8.164 & 0.703 & 0.728 & 0.740 & 0.764\\
\hline
LMC X-4 & 1.04 & 7.180 & 7.287 & 7.378 & 7.453 & 0.466 & 0.478 & 0.489 & 0.499\\
\hline
\end{tabular}}
\end{table*}

\begin{table*}[thbp]
\centering
\caption{{ Predicted radii and MI for some compact stars for different values of $\omega$ with $\alpha=0.596,\beta=-0.010888,\Phi_0=0.6$ correspond to Fig. \ref{f9}.}}\label{tab5}
 \scalebox{0.95}{\begin{tabular}{| *{10}{c|} }
\hline
{Objects} & {${M \over M_\odot}$}   & \multicolumn{4}{c|}{{Predicted $R$ km}} & \multicolumn{4}{c|}{{$I \times 10^{45} \,g-cm^2$}}\\
\cline{3-10}
&& \multicolumn{4}{c|}{$\omega_{BD}$} & \multicolumn{4}{c|}{$\alpha$}\\
\cline{3-10}
&  & 20 & 15 & 10 & 5 & 20 & 15 & 10 & 5  \\ \hline
PSR J1903+327   &   1.667  &  8.626  &   8.881  &   9.120  &  9.343  &   1.113  &   1.183  &   1.253 & 1.310 \\
\hline
Cen X-3 & 1.49 & 8.339 & 8.8562 & 8.752 & 8.977 & 0.923 & 0.970 & 1.015 & 1.062\\
\hline
EXO 1785-248 & 1.30 & 7.973 & 8.132 & 8.292 & 8.483 & 0.732 & 0.758 & 0.792 & 0.821\\
\hline
LMC X-4 & 1.04 & 7.288 & 7.495 & 7.591 & 7.734 & 0.480 & 0.498 & 0.516 & 0.536\\
\hline
\end{tabular}}
\end{table*}
\begin{table*}[thbp]
\centering
\caption{{Effects of $\alpha$ and $\omega_{BD}$ on mass, radius and moment of inertia corresponds to Fig. \ref{f9}.}}\label{tab6}
\begin{tabular}{| *{6}{c|} }
\hline
$\alpha$ & $\omega_{BD}$  & $M_{max}/M_\odot$  & $R$ km & $I \times 10^{45} \,g-cm^2$ &  $\mathcal{B}$ MeV / fm$^3$\\
\hline 
{0.397}  &   20  &  1.915  &   8.406  &  1.285 &   {56.998}  \\
\cline{2-5}
& 15  &   1.985  &  8.75 &   1.431 &     \\
\cline{2-5}
& 10  &   2.051  &   9.027  &   1.583 &     \\
\cline{2-5}
& 5  &   2.120 &   9.340  &  1.746 &   \\
\hline
{0.596}   &   20  &   1.940  &   8.550  &  1.375 &   {79.237}\\
\cline{2-5}
& 15  &  2.022 &  9.084  &   1.577 &    \\
\cline{2-5}
& 10  &  2.114  &  9.568  &   1.087 &    \\
\cline{2-5}
& 5  &   2.178  &   10.050  &   2.042 &    \\
\hline
\end{tabular}
\end{table*}

\section{Discussion and conclusion}

\begin{figure*}[thbp]
\centering
\includegraphics[width=8cm]{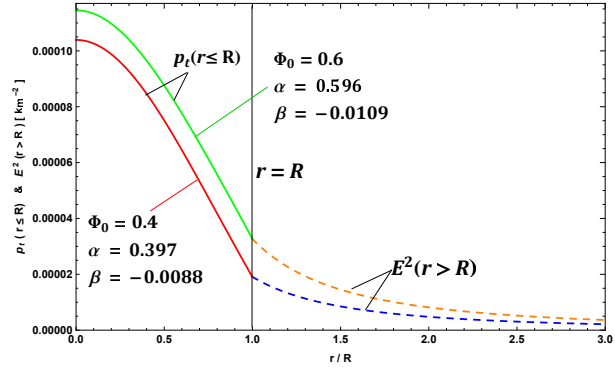}
\caption{Matching of $p_t(r \leq R)$ with $E^2(r>R)$ at the surface of the star.}\label{f10}
\end{figure*}

It is clear from the graphical analyses of our solution that the model of charged anisotropic strange star within the framework of Brans-Dicke gravity with a massive scalar field describes realistic stellar objects. The stellar model presented here obeys all the conditions required for hydrostatic equilibrium, stability and causality. Of particular interest is the contribution of the scalar field to the thermodynamical and gravitational properties of the stellar model. In Fig. 1 (right panel) we observe that an increase in $\alpha$ which corresponds to a larger scalar field intensity leads to higher densities. This observation supports the fact that the radial and tangential stresses also increase as $\alpha$ increases. The anisotropy parameter is also strengthened in the presence of larger scalar fields. Since $\Delta > 0$ throughout the stellar configuration, the repulsive force due to anisotropy helps stabilise the more compact configurations. An interesting observation is the increase in electric field intensity with an increase in scalar field intensity. Although these fields emanate from totally different sources there appears to be a `coupling' which manifests in the formation of more compact stellar configurations. It can also be seen from Fig. \ref{f10} that the tangential pressure at the interior of the stellar system continues as electric field at the exterior. This was also mentioned by Boonserm et al. \cite{boon} that for $p_t > p_r$ the scalar field satisfies $|\nabla \Phi|=0$ and the scalar charge density vanishes. This leads to the  non-vanishing electric field in such a region which falls off as $1/r^2$.

It has been shown that higher order gravity theories predict more compact objects compared to their 4D counterparts. The compactification is attributed to higher dimensional effects rather than exotic matter states such as dark matter. Our model provides an alternative mechanism for the existence of more compact objects than their classical relativistic counterparts. Recent studies of observational data obtained via gravitational redshifts have determined the the mass-radius relation of white dwarfs to a higher degree of accuracy. These results help constrain the equation of state of these compact objects thus giving us an handle on the matter composition and microphysics at play within the stellar fluid. Tables 1-3 display the upper bound and lower bound limits imposed by the modified Buchdahl limit for charged compact objects in Brans-Dicke theory. \\
In Tables  1 \& 2, we have generated values for the surface charge for well-known compact objects PSR J1903+327; Cen X-3; EXO 1785-248 \& LMC X-4 when $\Phi_0 = 0.4$ and $\Phi_0 = 0.6$ respectively with $\omega_{BD} = 5$. To see the effect of the BD parameter we have included calculated values for $\Phi_0 = 0.6$ but a much higher value $\omega_{BD} = 50$. It is clear from the data that an increase in $\Phi_0$ is accompanied by an increase in the surface charge as well as surface charge density. The increase in surface charge density is higher in more compact objects. This is expected as the charge contributes to the overall mass of the stellar body. We observe that the upper and lower bounds arising from the Buchdahl limit are modified by a change in $\Phi_0$. Table 1 shows that a decrease in compactness of approximately 15\% (from PSR J1903+327 to LMC X-4) is accompanied by a change in the lower bound as high as 40\% for $\Phi_0 = 0.4$  while the upper bound changes by approximately 1\%. An inspection of Table 2 reveals that a decrease in compactness of 18\% results in a decrease in the lower bound of approximately 65\% and a 6\% decrease in the upper bound.  In Table 3. we present model characteristics for $\Phi_0 = 0.6$ and $\omega_{BD} = 50$. We observe a decrease in 20\% in compactness is accompanied by a 60\% decrease in the lower bound and a 5\% decrease in the upper bound of the stellar configurations. A comparison of Tables 2 and 3 show the contributions attributed to the Brans-Dicke modification to the classical Einstein gravity theory,i.e. a change in $\omega_{BD}$. It is clear that  surface charge density, lower and upper bounds imposed by the Buchdhal limits are all affected by an increase in the Brans-Dicke coupling constant. Let us now turn our attention to the surface redshift for the different parameter sets displayed in Tables 1-3. It is clear that the surface redshift decreases as the compactness decreases.  We observe that surface redshift values obtained in Tables 1-3 for the stellar objects displayed here are consistent with the acceptable upper bound for for relativistic stars ($Z < 5.211$) \cite{ivanz}.
Another well-known characteristic named as the static stability criterion or $M-\rho_c$ function plays a crucial role in ensuring the stability of spherically symmetric static celestial systems under radial pulsation has been well-satisfied. We can also notice from the data drawn in $M-\rho_c$ curve that the celestial configurations become more massive according to increasing central density.

Further, we tested the state of the compact celestial bodies as well as the efficacy and the sensitivity to the stiffness of an EoS by studying the $M-R$ and $M-I$ diagrams generated from our celestial model. From $M-R$ curves that due to $\Phi_0$, $\omega_{BD}$ and $\mathcal{B}$ we envisaged two cases. The first case corresponds to $\omega_{BD}$ by setting $\Phi_0$ and $\mathcal{B}$ to $0.4$ and $56.998~MeV/fm^3$ respectively, one can see that when the parameter $\omega_{BD}$ decreases from $20$ to $5$, the maximum value of mass $M$ increases with the increasing radius $R$, which produces more massive compact celestial bodies. The second case corresponds also to $\omega_{BD}$ by setting $(\Phi_0,\mathcal{B})=(0.6,79.237)$, we can observe that when $\omega_{BD}$ increases from $5$ to $20$, the maximum value of mass $M$ and corresponding $R$ decreases, which gives us a celestial system less compact and less massive for increasing $\omega_{BD}$ with fix $\Phi_0$ and $\mathcal{B}$. Moreover, from $M-I$ curves, we can see that the maximum moment of inertia $I$ is always increasing with an increase in the mass until up to the maximum value of mass and decreasing rapidly with decreasing the mass under the effect of $\Phi_0$, $\omega_{BD}$ and $\mathcal{B}$. In this respect, we can conclude that the stiffness of EoS is better in the case of $\Phi_0=0.6$ and $\omega_{BD} =5$ while compared to all other cases, i.e., when $\omega_{BD} =10,15,20$ and $\Phi_0 =0.4,0.6$. We also found a good agreement with observational data on $M-R$ and $M-I$ diagrams for four compact celestial bodies viz., PSR J1903+327, Cen X-3, EXO 1785-248, LMC X-4 and many others can be adapted. The tables (4)-(6) display the values of physical parameters such as maximum mass, radius and momentum of inertia corresponding the Fig. (\ref{f8}) and (\ref{f9}) for different values of $\Phi_0$, $\omega_{BD}$, and $\mathcal{B}$. With the above rigorous analyses of the gravitational and thermodynamical behaviour of our solution we ascertain that our model meets the necessary requirements for a physically realizable self-gravitating compact object.

In this study we have generated a model of a compact charged stellar object within  the Brans-Dicke gravity framework in the presence of a massive scalar field. In addition, the matter composition of the stellar interior obeys the MIT Bag model equation of state. Our model satisfies all the criteria for a genre of compact objects which include strange stars. The highlight of our work is the interplay between the electric and scalar fields which originate from completely different sources combine to affect physical characteristics of the model. The Brans-Dicke coupling constant also affects stellar characteristics such as compactness, redshift and the bounds required by the modified Buchdahl limit for charged stars. In addition, we observed that  $M-R$ and $M-I$ curves are sensitive to changes in $\Phi_0$, $\omega_{BD}$ and the Bag constant which in turn points to a change in stiffness of the stellar fluid. We believe that this is a novel feature in our model which inherently connects the macrophysics (scalar field, electric field and BD coupling constant to the microphysics (Bag constant). It would be interesting to compare and contrast our findings to phenomenological features derived in higher dimensional gravity theories such as Einstein-Gauss-Bonnet gravity.

\end{document}